# What do we learn from impurities and disorder in High $T_c$ cuprates?

Henri ALLOUL

Université Paris-Saclay, CNRS, Laboratoire de Physique des Solides, 91405, Orsay, France

## Abstract

We have performed a series of experimental studies that established that the differing morphologies of the **phase diagrams** versus hole doping $n_h$ of the various cuprate families are mostly controlled by defects and disorder. YBCO being among the minimally disordered cuprates we could introduce controlled defects allowing us to probe the metallic and superconducting states. We could demonstrate that the extent of the spin glass phase and the superconducting dome can be controlled by the concentration of spinless (Zn, Li) impurities substituted on the planar Cu sites. NMR frequency shift measurements establish that these defects induce in their vicinity **a cloud with a Kondo like paramagnetic behavior.** Its "Kondo" temperature and spatial extent differ markedly between the pseudogap and strange metal regimes. We have performed transport measurements on single crystals with a controlled content of in plane vacancies introduced by electron irradiation. At high $T$ the inelastic scattering of the carriers has been found independent of disorder and completely governed by the excitations of the correlated electronic state. The low $T$ upturns in the resistivity associated with single site Kondo like scattering are qualitatively in agreement with the local magnetism induced by spinless impurities. The apparent metal insulator crossover are only detected for very large defect content and part of the large resistivity upturn remains connected with the Kondo like paramagnetism. In the **superconducting state**, the defect induced reduction of $T_c$ scales linearly with the increase of residual resistivity induced by disorder. High field magnetoresistance experiments permit us to determine the paraconductivity due to superconducting fluctuations. The latter vanishes beyond a temperature $T'_c$ and a field $H'_c$ that decrease with increasing the in plane defect content. In the pseudogap regime the weaker decrease of $T'_c$ with respect to that of $T_c$ reveals a large loss of superconducting phase coherence in presence of disorder. In the light of our experimental results, we initiate a discussion of its interplay with pair breaking.  Our data also permit us to confirm that the differing phase diagrams are due to competing orders or disorders that are family specific. In the ideal phase diagram of a disorder free cuprate the 2D superconductivity should persist at low doping. This ensemble of experimental results provides serious challenges for the theoretical understanding of superconductivity in these correlated electron systems.

## (A) INTRODUCTION

Impurities and disorder play a major role in Solid State Physics as could be already evidenced in materials for which an independent electron approximation holds. They govern the mechanical and optical properties of most materials. The chemical nature of impurities determines for instance the beautiful colors of gems that are large bandgap insulators. The transport properties of semiconductors were only under control when it became clear that the carrier content is associated with the chemical properties of the impurities and their concentration. The low temperature electronic and thermal conductivity of metals are limited by the scattering of carriers on impurities and lattice defects. In the metals that display a superconducting state at low $T$ the disorder favors the type II state and plays a major role in the pinning of vortices, and so on…

One does naturally expect then to find that disorder has a strong incidence on the original physical properties displayed by materials in which electronic interactions are dominant, such as the high $T_c$ cuprates (1). Those are materials involving a large number of atomic species, and are then akin to house a variety of impurities or lattice defects. On the experimental side they are quasi unavoidable and intrinsic to the growth of the materials. The basic structure of cuprates is a set of layered $CuO_2$ planes separated by charge reservoir layers. The former are responsible for the original physical properties, while the second are essential to control the electron or hole doping of the $CuO_2$ planes. On the theory side, one faces a sufficiently complex theoretical challenge to try to explain the properties of a single $CuO_2$ plane with increasing doping. One generally considers impurities and disorder as unwanted complications.

In actual cuprates defects are always present and, for experimentalists, a real difficulty is to decide whether the observed physical properties are generic, that is due to the ideal defect free cuprate plane. Most of the work reported in this article are attempts to isolate those from the extrinsic effects related with the chemical doping procedure. I shall altogether show as well that defects or impurities injected in a controlled manner can be excellent probes of the properties of the pure material.

An essential approach to reach this aim is to eliminate as much as one can the sources of disorder that smear or influence the measured properties. Improving the sample preparation procedure is an obvious method to achieve this task. An alternative approach is to compare the properties of different cuprate families to sort out the generic properties from those that are specific or marginal. Once this has been performed, that is a rather disorder free compound or family has been selected, one might then try to introduce defects in a controlled manner in that pure material. One may then use both thermodynamic and local probes to determine the incidence of specific defects on the physical properties. I shall give evidence that those observations sometimes bring new information hardly accessible experimentally on the pure material (2).

I have initiated such an orientation in my research group since the early days of the discovery of high $T_c$ superconductivity and encouraged many researchers in the Orsay-Saclay area to follow this approach over the last 35 years. I summarize here the various important results obtained that allowed some understanding of the experimental properties of cuprates and still raise important questions to be resolved on the theory side.

I shall consider first in section **B** the $(T,n_h)$ phase diagram of diverse hole doped cuprate families and give evidence that its morphology is driven by the existing disorder, which raises the question about the actual phase diagram which should represent the hypothetical disorder free cuprate. This will further allow me to categorize the cuprate families with increasing disorder. I shall then show that the YBCO family is among the cleanest cases where one could investigate the incidence of an increasing content of defects. This allowed me to present in section **C** the local NMR studies of the magnetic properties induced by Zn and Li spinless impurities in the metallic state of the $CuO_2$ planes. Using highly energetic electron irradiation allowed one to control the introduction of vacancies in the $CuO_2$ planes in cuprate single crystals. Their incidence on the transport properties were identical to that produced by Zn chemical substitutions. We could therefore perform a refined comparison of the evolution with defect content of the normal state transport and magnetic properties. In section **D** I shall then present results on the corresponding modifications of the SC properties such as the decrease of $T_c$. Similarly, I shall describe a detailed investigation of the evolution with disorder of the superconducting fluctuations range, using high magnetic fields when necessary. The importance of the loss of phase coherence in the 2D SC state in presence of disorder will be underlined. A discussion of this ensemble of results, done in section **E,** will permit to summarize that one cannot any more consider disorder effects as unnecessary complications. They rather permit to raise many essential questions and give useful guidance for theory.

## (B) CUPRATE PHASE DIAGRAMS AND DISORDER

Let me perform first a small historical survey about the discovery of the cuprates and of their phase diagram and discuss hereafter my present understanding of the incidence of disorder.

### *1) Early phase diagrams of the cuprates*

The first cuprate discovered by Bendorz and Muller has been the family of 124 compounds (3). The parent $La_2CuO_4$ is an AF Mott insulator with $T_N$=240K that can be doped by substitution of a concentration $x$ of $La^{3+}$ by $Ba^{2+}$ or $Sr^{2+}$ (4). The long range ordered AF state is destroyed for $x$=2 % and a SC dome appears for $0.07 < x < 0.3$ with a maximum $T_c$= 38K for $x$~0.19. From simple chemical arguments the concentration of holes in the $CuO_2$ planes is merely $n_h=x$ and most researchers have immediately considered that this $(T,n_h)$ phase diagram is generic of the cuprate plane. One noticed as well that the material displays in the intermediate range 0.02<x<.0.07 a low $T$ disordered frozen spin state of the Cu $3d^9$ magnetic sites labelled as spin glass (SG).

A year later the $YBa_2Cu_3O_{6+x}$ family has been discovered (5), with also a parent AF state with $T_N$=410K for $x$=0 and part of a SC dome with a maximum $T_c$ = 93K. This compound involves two $CuO_2$ layers with an intermediate $Y^{3+}$ layer. Successive bilayers are separated by a square lattice layer of non-magnetic $Cu^+$ for $x$=0. The hole doping for $x$>0 results from the extra oxygen atoms introduced in this intermediate Cu layer. The relation between $x$ and $n_h$ is not straightforward as the $Cu^{2+}$-$O^{2-}$- $Cu^{2+}$ chain segments formed do not transfer charge in the $CuO_2$ planes. The phase diagram displayed in Fig. 1 is quite similar to that of $La_2CuO_4$, but the AF phase is contiguous to the SC phase (6). An important difference is therefore the absence of intermediate SG regime. While this is a very important distinction, many researchers mostly interested by the SC properties considered that as a minor difference and mapped the $n_h$ ranges in the 124 and YBCO phase diagrams by an artificial parabolic fit of the SC domes (7).

In our preliminary experiments with Zn impurities in $YBa_2Cu_3O_{6+x}$ we immediately found that the reduction of $T_c$ was larger in the underdoped regime than in the optimally doped case. This Zn induced in plane disorder has then led us to notice changes of the morphology of the phase diagram (Ph. D.), as displayed in Fig. 1 (8). There we plotted together the Ph. D. versus oxygen content *x* for pure samples and for 4% Zn substitution (9). One can see that both extents of the AF phase and the SC dome shrink and disclose a wide SG regime. The morphology of the Ph.D. resembles then markedly that of the 124 family upon this introduction of disorder. The latter plays therefore a major role in defining the extent of the SG regime in the cuprates. This observation also implied that the $n_h$ mapping between phase diagrams is not accurate.

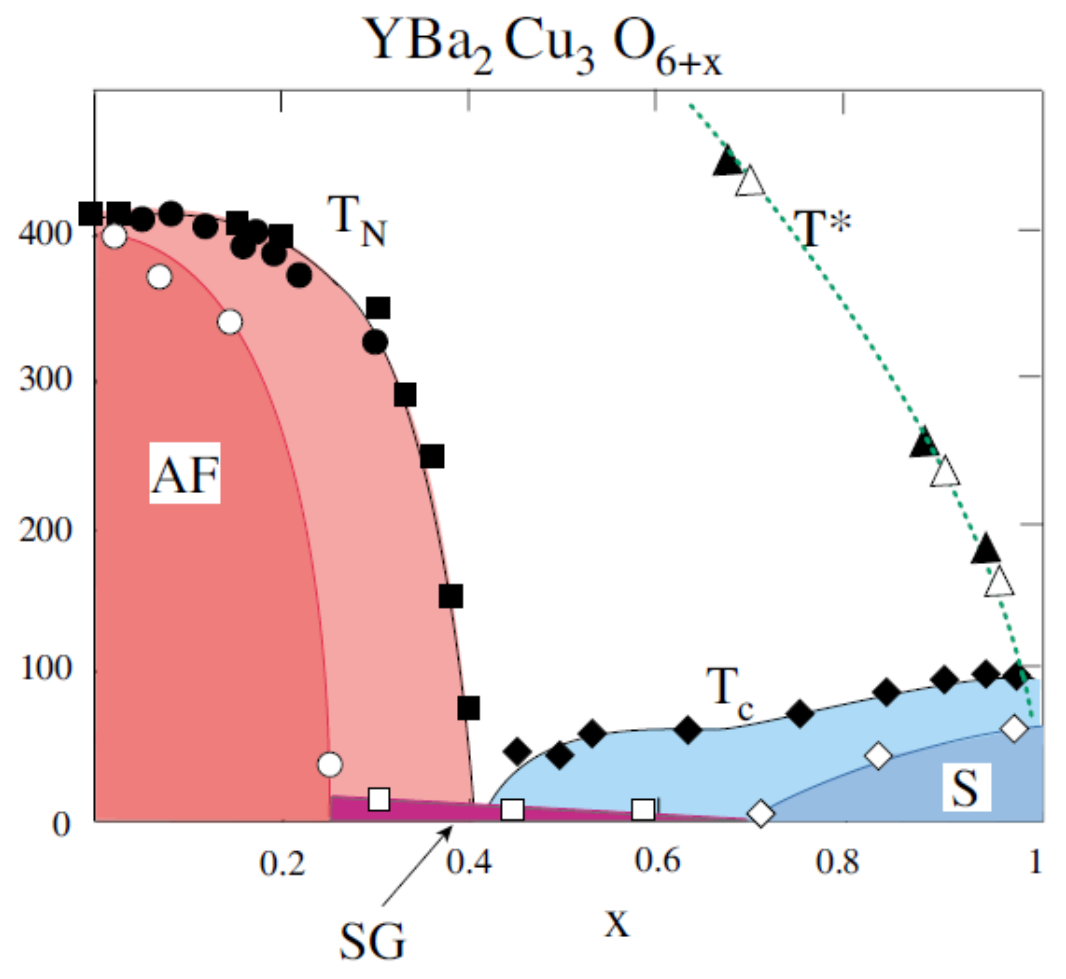

Fig 1 Phase diagrams of the AF Néel temperature $T_N$, the Spin Glass transition $T_{SG}$, the SC transition $T_c$ and the pseudogap $T^*$ for pure YBCO samples (filled symbols) and for 4% Zn substitution (empty symbols). From ref. (9).

That occurred at the very time of the PG discovery in the underdoped pure YBCO using $^{89}$Y NMR shift data (10). I initiated these Zn substitution experiments mostly to try to understand whether the PG had any relation with superconductivity. With 4% Zn substitution our $^{89}$Y NMR data immediately convinced us that *T** is identical for the underdoped $YBCO_{6.6}$ sample for which SC is fully suppressed (8). This total independence of the two effects at large distance from the impurities suggested that the PG is a phenomenon that remains robust at short range and competes with the SC state. This powerful indication revealed by introduction of Zn impurities does not however allow to conclude whether *T** corresponds to a phase transition in the defect free compound. The actual disorder introduced by the Zn substitution might indeed prevent a long range ordering.

We were at that time most concerned by an important question concerning the PG. What could explain the large $T_c$ difference at optimal doping between the 124 and YBCO families? Could that be associated with an intrinsic difference between compounds with single $CuO_2$ layers and those with bilayers in the unit cell? We completely cleared that question by performing the first study of powder samples of the single layer Hg1201 cuprate family for which the optimum $T_c$ is about 85K. We could synthesize samples covering a large range of doping from the underdoped to the overdoped regime. The comparison of $^{17}$O NMR shift data with that taken in the YBCO family allowed us to demonstrate that the PG *T** are in all respects quantitatively identical in both families (11).

In this Hg cuprate family the doping proceeds again from an increase of O content in the Hg layer that is far from the $CuO_2$ plane, so that the dopant induced disorder should be weaker than in the 124 family. These experiments confirmed therefore that the PG is a robust phenomenon independent not only on the number of $CuO_2$ layers but also on the disorder induced by dopants or substitutions.

## 2) A 3D phase diagram

The Zn substitution experiments of Fig. 1 indicate that the SC dome shrinks with increasing in plane disorder. This is evidenced as well using electron irradiation to increase progressively the in plane defect content (see section **C**). These experiments have led us altogether to understand that the morphology of the Ph.D evolves with the increase of disorder induced by the increase of Zn content. We could then propose the 3D phase diagram of Fig.2 in which the third axis represents the increase of Zn content (12). Transport experiments to be detailed in section **C** allowed to follow the iso-$n_h$ red lines plotted in Fig.2 that confirm the average displacement of the SC dome to the right in the Ph.D while the SG range broadens and the PG $T^*$ is unchanged.

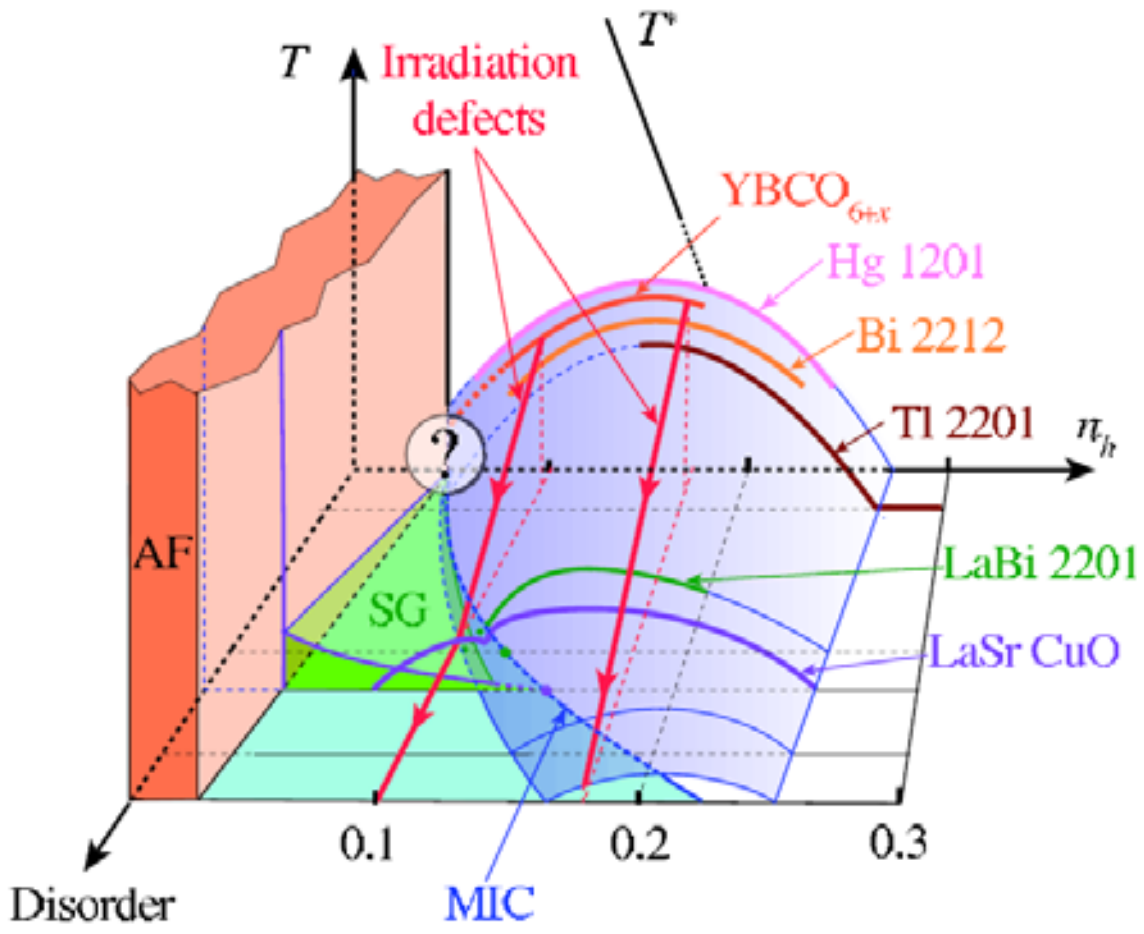

Fig. 2 Variation of the ($T$,$n_h$) phase diagram for increasing disorder. The red lines represent the $T_c$ variation with disorder induced by electron irradiation in YBCO (see section **C**). The doping range of the AF state slightly shrinks with increasing disorder for instance with Zn content, or with Ca content in $Y_{1-y}$ $Ca_yBCO_6$ (see text in **B**). The $T_c$ domes reported for the various cuprate families are limited to the experimentally accessible doping ranges. Their position on the disorder axis are chosen in order to roughly match their SG ranges with those expected with increasing Zn induced disorder in YBCO. From ref. (12).

We shall consider hereafter the various cuprate families and shall try to anticipate where their phase diagram could be located on such a 3D scheme. We discarded there the Ph.D for homovalent substitutions of Rare Earths on the $Y^{3+}$ sites in $YBCO_{6+x}$ as they do not significantly modify $T_c$ and the Ph.D. We did not consider either the peculiar case of Pr, that was shown to display a 4+ valence that changes the oxygen content and hole doping for both AF and SC (13) (14). For heterovalent substitutions one can assume that the disorder due to the coulomb interaction with the dopants decreases as their distance to the $CuO_2$ planes increases.

For all the single layer or bilayer cuprate families the PG $T^*$ variation observed by various experimental techniques were found to match with those observed for YBCO. For those families that display $T_c$~90K at optimal doping the dopants are located in the interlayers. So for these families the mapping of the ($T_c$,$n_h$) SC domes usually assumed by parabolic fits might be acceptable. In Fig. 2 their SC domes are located very near that of YBCO.

Let me point out that a weak dopant induced disorder remains in those actual compounds for which the optimum $T_c$ is about 90K. The AF and SG domains for those families could not be probed except for YBCO as they could not be sufficiently underdoped experimentally. We cannot say at this stage whether in absence of disorder the $T_c$ dome would reach $T_c$ =0 before the onset of the AF state. Would the clean phase diagram display a first order transition from the AF to the SC state? In the Fig.2 we introduced therefore the question mark near the 2D phase diagram for vanishing disorder.

The heterovalent substitutions in (LaSr)124 and (La-Bi)2201 families occur on sites adjacent to the $CuO_2$ planes. We assumed that the disorder is large in these systems and we consequently located their 2D Ph.D along the “disorder” axis at points for which their SG ranges approximately match those for YBCO+Zn.

We also considered another Ph.D. for $Y_{1-y}Ca_yBCO_6$ in which the $n_h$ doping proceeds from a partial substitution of $Ca^{2+}$ on the $Y^{3+}$ site inserted between the $CuO_2$ layers in the parent $YBCO_6$ (15). We again found that the AF state disappears for a Ca content $y=2n_h$ of about 7%. A SG range was disclosed for $0.03< n_h <0.09$ and a maximum $T_c$ of about 30K is reached for $y=0.25$ that unfortunately appears to be the Ca chemical solubility limit. Here the large disorder induced by the substitution site between the $CuO_2$ planes again generates a Ph.D that maps even quantitatively that of (LaSr)124 for $n_h <0.15$. We did not plot it in Fig. 2 to keep the figure clear.

In our proposed 3D Ph.D. the quantitative definition of the disorder for a given family is of course rather rough. The disorder magnitude depends of the defect types and even evolves with doping inside many cuprate families. The data for a given family should be on a distorted surface in Fig.2 rather than on parallel flat planes.

The 3D representation however gives a reliable indication on the actual trend followed in the real materials. The systematic analysis done allows us to conclude that the disorder governs the morphology of the ($T$, $n_h$) phase diagram by opening the spin glass regime.

This 3D phase diagram even suggests that the disorder governs the optimal $T_c$ value of the diverse cuprate families. In plane disorder produced by Zn lead to a subsequent decrease of optimum $T_c$. Out of plane disorder has also an incidence on $T_c$ values, as illustrated for instance within the 124 family. There $T_c$~45K are obtained by electrochemical intercalation of excess oxygen (16) that induces less disorder than $Sr^{2+}$ substitution on the $La^{3+}$ sites.

From the above phase diagrams analysis we cannot however conclude whether the difference of optimal $T_c$ is solely due to the existing disorder. Further data will allow us to discuss again that point later in section **E**.

Here I did not discuss so far the interesting low $T$ phases detected well below $T^*$ that have been thoroughly studied during the last decades. Soon after the cuprate discovery a stripe charge order has been highlighted in the 124 phase (17). It is strongly pinned for $Ba^{2+}$ substitution on the $La^{3+}$ site and fully depresses the SC state for $x$=0.12. For $Sr^{2+}$ substitution this stripe phase is less pinned and only seen as a small dip in the $T_c(n_h)$ variation.

In YBCO many unexpected phenomena were detected at low $T$ in high quality samples with $x$~0.6, starting by Quantum Oscillations on the resistivity (18), corresponding to a small Fermi Surface orbit and a negative Hall Effect (19). Then NMR in high applied field allowed to evidence a charge ordering below $T_c$ =60K (20). The latter could possibly be associated with the 2D CDW order detected for $T$ as high as 150K by RIXS experiments in zero field (21). Extension of these measurements in high field (22) gave final evidence that the 2D CDW transforms below 60K into the 3D CDW seen by NMR. The present understanding is that a reconstruction of the Fermi Surface occurs below the SC dome, in a narrower dome shaped region of the phase diagram (19) centered at about x=0.6. This original coexistence of SC and CDW order parameters has been underlined and has justified extensive theoretical studies (23) (24).

Independently of the originality of this situation I anticipated early on that these electronic orders occurred well below $T^*$ and therefore could not be responsible for the pseudogap that is highly related to Mott physics (25). Nowadays an emerging consensus is that the cuprates can display diverse low $T$ electronic orders in the PG regime (19). Those compete with SC and are somewhat distinct in the diverse cuprate families, as found in 124 and $YBCO_{6.6}$

I have always suggested that the actual disorder or order introduced by the hole dopants plays a great role in selecting the ground states (25). The CDW dome in YBCO occurs for a hole doping nearby that for which a slight singularity occurs in the $T_c(x)$ variation of

Fig. 1. Structural studies have also evidenced that in this range of oxygen contents in YBCO a Cu-O chain ordering occurs in the interlayer Cu plane (26). I am however not aware of any careful experimental effort to search whether some correlation holds between the Oxygen structural order and he CDW. In any case I sill suggest that these competing charge orders are therefore not generic of the ideal $CuO_2$ plane dreamed of by theoreticians.

## (C) POINT DEFECTS IN THE PSEUDOGAPPED AND STRANGE METALS

The *d* wave superconductivity, the PG and the strange metal with a *T* linear behavior of the resistivity for optimal doping are the actual generic physical properties of the cuprates still requiring a thorough understanding. Experimentally one still needs to gather all possible information on the cleanest materials, which from the above analysis should be among the families with the highest optimal $T_c$.

The family of $YBCO_{6+x}$ appears to be a favorable case as the full range from the AF to the slightly overdoped regime is accessible without the occurrence of the intermediate SG phase. This indicates that disorder effects are minimal as we anticipated from NMR studies of the various nuclei of the unit cell (27).

For some specific oxygen content the disorder might even be rather weak. The AF state for $x$=0 is undoped with a half filled Cu level in the $CuO_2$ planes and is not modified by a weak excess oxygen content. For $x$=1 the unit cell is stoichiometric with filled ordered chains. It seats in the slightly overdoped regime with $T_c$=92K, about 1K below that at optimal doping. In the PG regime we found out that the sample reproducibility is extremely good with a large $T^*$~350K for $x$=0.6. This slightly exceeds the composition 0.5 with alternative full and empty chains but one anticipates again that the excess $Cu^{2+}$-$O^{2-}$-$Cu^{2+}$ chain segments do not contribute significantly to the charge transfer, hence the flat part of the $T_c(x)$ curve in the phase diagram in Fig. 1. We have subsequently considered for years that these three oxygen contents were among the purest references for the AF, the strange metal and the large PG metal phases of the cuprates. The highly overdoped regime has been mostly accessible in the Tl2201 family (28). We concluded therefore for long that one might then use these starting materials to introduce controlled defects and study in some detail their incidence on the physical properties.

Why should we introduce defects? In classical physics, when we perfectly know the properties of a material we might anticipate the incidence of the defects introduced. For instance if one drops a stone on a pond of a given liquid, one knows that a set of waves is created and propagates as circles on the surface. Their period and damping tells something about the viscosity of the liquid of the pond. Similarly introducing an impurity in a metal gives the Friedel charge density oscillations (29) or RKKY oscillations of spin density if the impurity is magnetic (30) (31). If one considers now a poorly understood correlated electron compound, one may not anticipate the disturbance created. Then the study of the response to the impurity is all the more important as it gives information on the properties on the pure material (2). This is what we have been doing for years in the cuprates. I shall summarize hereafter first in §1 the information on the magnetic properties gained mostly by NMR techniques using impurity substitutions. In §2 I shall present the incidence on the transport properties of impurities and vacancies in the $CuO_2$ planes. The latter, obtained in a controlled manner by electronic irradiation of single crystals allowed us to probe the properties of the distinct metallic regimes.

### *1) Magnetic properties induced by spinless impurities*

In the 2D cuprates the largest perturbations are induced of course by defects located in the $CuO_2$ planes. The substitution of $Cu^{2+}$ by $Zn^{2+}$ done from the early days of HTSC introduces a spin vacancy without any charge defect. In the AF state of $YBCO_6$ these spinless defects only suppress spins on the AF lattice. This spin dilution effect is sufficient to explain the reduction of $T_N$ with increasing Zn content (32).

For a given Zn content the introduction of Oxygen on the intermediate Cu plane leads to the same hole doping as for the pure material. Apart the appearance of a SG phase, we could immediately detect an increase of the $^{89}$Y NMR linewidth that increases with decreasing $T$ down to $T_c$. Such a broadening can only be associated with the appearance of local magnetic moments (8).

In that regime when the oxygen content is large enough the spin-glass temperature $T_{SG}$ becomes negligible above $x$=0.6 in Fig.1. Spinless impurities never induce such paramagnetic local states in standard independent electron metals. SQUID measurements of the spin susceptibility allowed a determination of an upper limit of its magnitude (33). An increase of the NMR experimental sensitivity enabled us then to reveal for x=0.6 weak $^{89}$Y NMR signals shifted from the main $^{89}$Y NMR line, with a $T$ dependent NMR shift. We could attribute those "satellite" lines to $^{89}$Y nuclei located near the Zn substitution site (34).

These paramagnetic moments appear therefore as a cloud of Cu sites with some spatial extension around the Zn site. They are characteristic of the correlated electronic nature of the pseudogap regime. In the slightly overdoped $YBCO_7$ samples a weak broadening of the Y NMR line has been detected but no satellite could be resolved in this strange metal regime above $T^*$ and $T_c$, so that the magnitude and/or the spatial extent of the moments is weaker/smaller.

We performed then an important step in order to test the eventual incidence of a charged defect. We used $Li^+$ substitution on the $Cu^{2+}$ site of the $CuO_2$ planes and surprisingly discovered that the $^{89}$Y NMR spectra were identical for 1% substitution of Li or Zn (35). Both types of defects induced the same paramagnetism in their vicinity. The extra electron given by the Li therefore just fills a hole state in the bath band. This only slightly modifies the overall hole content for dilute Li substitution. This experimental result allowed us to evidence that the dominant defect in the $CuO_2$ plane is the spin vacancy associated with the substituent rather than its local charge.

As this appeared awkward we performed a similar comparison in another correlated electron system, $Y_2BaNiO_5$ in which the $Ni^{2+}$ are ordered in chains of *S*=1 spins that constitute a prototype of Haldane chain (2). There the substitutions on the Ni site disrupt the chains, a staggered paramagnetism occurs near the chain ends and the $^{89}$Y NMR displays then many satellite lines (36) (37). Those were identical for Zn, Cu and Mg substitutions so that any disruption of the spin chain always induces the same staggered paramagnetism on the chain ends. The NMR data allowed a measure of the extension $\xi_{imp}$ of this staggered perturbation. Its *T* dependence accurately fits the theoretically computed magnetic correlation length $\xi$ for the pure chain.

Assuming a similar staggered local 2D paramagnetic cloud with an exponential correlation length for the cuprates, we have estimated that $\xi_{imp}$ would increase from about two to five lattice constants in the PG range from *T** to $T_c$ for *x*=0.6 (38). The intrinsic width of the $^{89}$Y NMR forbids any accurate fit with such a model for the *x*=1 strange metal, but $\xi_{imp}$ would not exceed two lattice constants and would be roughly *T* independent. Such a model dependent determination is rough but establishes a qualitative difference between the PG and strange metal regimes. These $\xi_{imp}(T)$ would also agree with evaluations of the magnetic correlation length $\xi$ done from inelastic neutron scattering experiments in single crystals of pure YBCO (39) (40). The staggered response detected around spinless impurities therefore monitors the actual correlation length of the pure cuprates, as found for the Haldane chain from both theory and experiments (2) (37).

We disclosed a further advantage of the Li substitution as the $^7$Li NMR signal magnitude and intrinsic resolution are larger than those for $^{64}$Zn. We could therefore detect its NMR shift $^7K$ for Li content as small as 1% whatever the hole doping in the metallic range of $YBCO_{6+x}$.

This NMR shift $^{7}K$ results from the transferred hyperfine coupling from the four Cu nearest neighbors of the Li in the $CuO_2$ plane. Those are the Cu sites exhibiting the largest paramagnetic response, in an exponentially staggered paramagnetic cloud model. We could measure accurately the $T$ variation of $^{7}K$ for any hole doping (35). As shown in Fig. 3a, it can be fitted quite accurately with a Curie Weiss dependence $^{7}K=C(T+\Theta)^{-1}$.

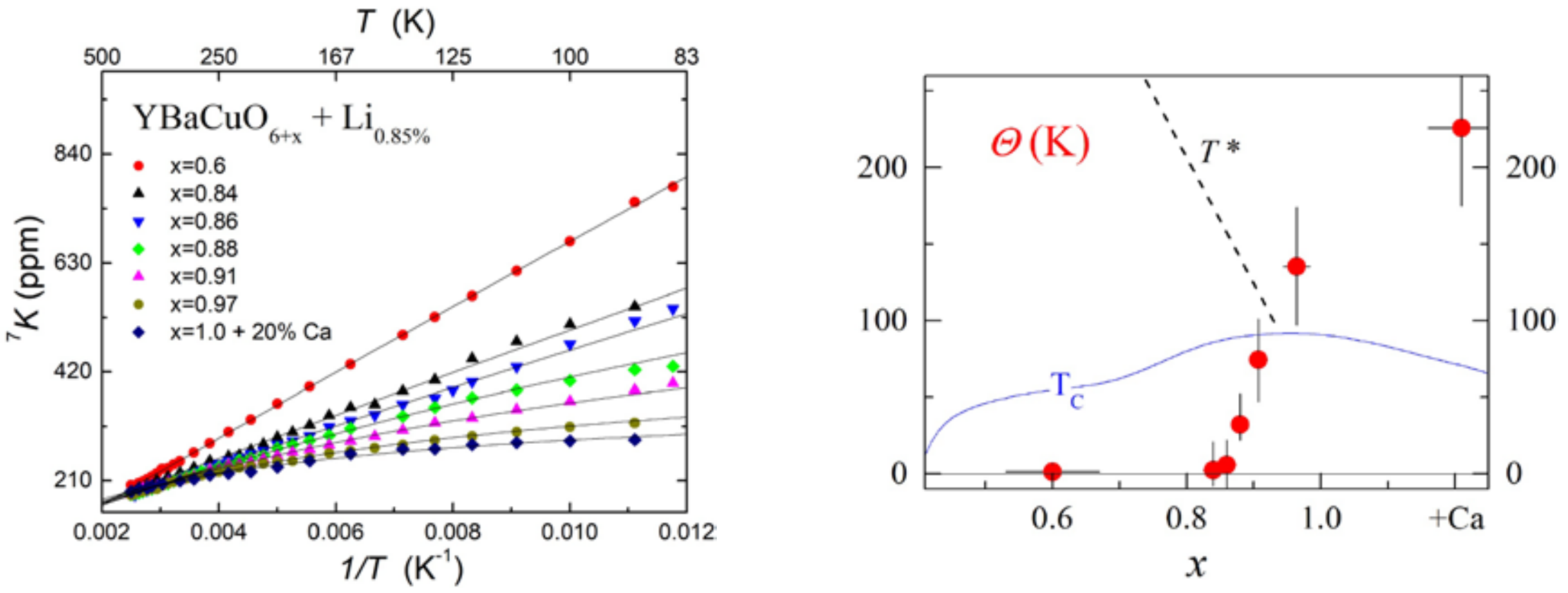

Fig. 3 **(a)** $^{7}$Li NMR shift $^{7}K$ plotted versus $1/T$ for increasing doping in $YBCO_{6+x}$, up to the overdoped regime with 20% Ca. The full lines are least square Curie Weiss fits. **(b)** Comparison of the $\Theta$ values deduced from the fits with the $T_c$ and $T^*$ data from Fig.1. From ref. (35).

The effective moment given by $C$ is found independent of hole doping within experimental accuracy. The Weiss temperature $\Theta$ shown in Fig. 3b exhibits a large change from the PG to the strange metal regime. While $\Theta$ is very small in the PG regime it abruptly increases through the PG "critical" doping point.

We did notice that this $T$ dependence of the local magnetic response is somewhat analogous to that known for Kondo magnetic impurities in classical Fermi liquid metals. The evolution with hole doping of the apparent Kondo temperature $\Theta$ is similar to that found when changing 3*d* impurities in pure copper metal for which the Kondo temperature $T_K$ changes from a few mK for Mn to 30K for Fe (41).

Let me point out that all these data of Fig. 3a were taken above $T_c$ and that for x=0.6 they do not show any singularity at 150K, the onset temperature of the 3D CDW detected by RIXS experiments (21). The variation of $\Theta$ with hole doping of Fig.3(b) is obvious even from the data taken above 150K, so that the magnetic response is not influenced by this charge order, if present in our samples. This response to Li impurities is therefore probing a specific property of the PG regime. Similar studies in other cuprate families have not been performed so far as intrinsic disorder due to chemical doping complicates the experimental situation. More material research efforts on Li substitutions in favorable cases such as BiSCO2212 or Hg1201 would be required.

I believe that this paramagnetic response to spinless impurity substitutions characterizes an abrupt evolution of the metallic state from the PG to the strange metal regime in the cuprates. It remains an important challenge for any theoretical interpretation of the correlated electronic properties of the cuprate physics. $^{17}O$ NMR experiments provide evidence that this paramagnetic behavior persists in the superconducting state of the PG phase, while the Kondo- like screening is quite reduced in the SC state of the strange metal (42).

### *2) Incidence of in plane defects on the normal state transport*

I shall consider hereafter the incidence of impurities and vacancies in the $CuO_2$ plane on the transport properties in $YBCO_{6+x}$ single crystals.

*(a) Experimental techniques*

F. Rullier-Albenque had been using homogeneous irradiation with a highly energetic (2Mev) electron beam to produce defects in classical metallic superconductors (43). She has extended her expertise to create vacancies in controlled content in thin cuprate crystals platelets. She checked first that the dominant defects acting on the planar transport are the vacancies produced in the $CuO_2$ planes by ejecting atoms out of these planes. A s comparison with initial work done with Zn substitutions established that both resistivity and Hall effect are similarly modified in these two cases (44) (45).

I present in Fig.4 clear evidence for the important advantages of this electron irradiation technique. There the resistivity curves *ρ(T)* are taken on $YBCO_7$ (46), $YBCO_{6.6}$ and Tl2201 (47) samples in which the damage created by electron irradiation is progressively increased. One notices immediately on these *ρ(T)* data some important features that will be discussed at length hereafter:

(1) The data accurately parallel each other at high *T.*

(2) Low *T* upturns appear with increasing damage. They suggest Metal Insulator Crossovers (MIC).

(3) The SC transition can be suppressed by disorder down to $T_c$= 0.

(4) A downturn of *ρ(T)* onsets above $T_c$ at a temperature $T'_c$ that signals the extension of the Superconducting regime.

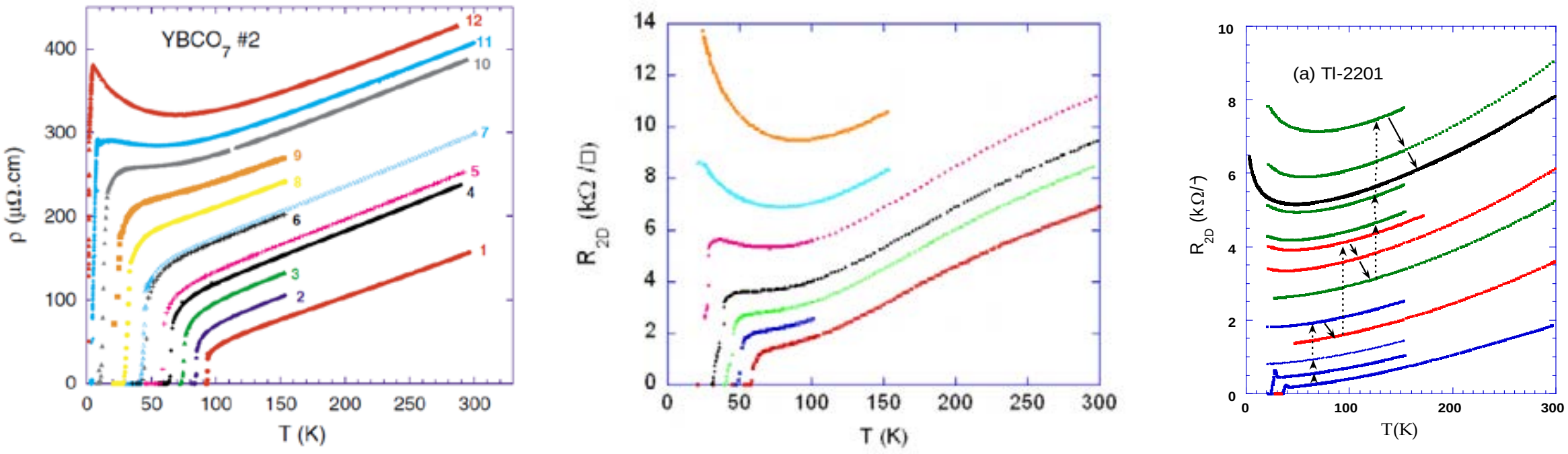

Fig.4: Planar resistivity data taken **(a)** on a virgin $YBCO_7$ (curve 1), and after steps of in situ electron irradiation at 20K. The curves from 2 to 12 are data taken while cooling after annealing at 150K or 300K. **(b)** on a $YBCO_{6.6}$ sample. The two upper data curves correspond to special heat treatments (see text). **(c)** on an overdoped sample of Tl2201 with $T_c$=31K. The arrows trace the thermal path followed during the experiment. The final black curve is a record after a long room *T* annealing. Figures respectively from ref. (46) and (47).

I shall analyze first in some detail hereafter points (1) and (2) related with the normal state properties. Experiments performed with large pulse field did allow us to suppress fully the superconducting Fluctuations (SCF) and follow the variation of the normal state resistivity below $T_c$. This also allowed then to evaluate the contribution of the SCF to the conductivity as described in section **D**.

The 12 $\rho(T)$ curves reported on of Fig. 4a had been measured in situ on a single $YBCO_7$ crystal within the cryostat where electron irradiation was performed at 20K. After recording a first cooling curve, $\rho(T)$ was measured after each irradiation step. The defects created at 20K are stable up to 150K and the resistivity curves are reversible, but heating to room *T* produces a partial annealing of some vacancies. The procedure was similar for the $YBCO_{6.6}$ sample of Fig. 4b. Before recording the upper data curve the sample was annealed at 400K to achieve a clustering of the defects and a new irradiation was performed to fully suppress superconductivity. For the overdoped Tl2201 sample of Fig. 4c the last curve has been taken in a distinct cryostat after a long room *T* annealing.

As will be detailed in section **D** high fields allow us to suppress the SC contributions (3) and (4) to the conductivity, once $T_c$ has been highly reduced for large enough defect content. We could then separate three contributions to the normal state resistivity

$$\rho_n(T) = \rho_0 + \Delta\rho(T) + \rho_{pn}(T). \qquad [1]$$

Here $\rho_{pn}(T)$ is the normal state resistivity measured for the pure sample. Both the *T* independent increase of “residual” resistivity $\rho_0$ and the low *T* upturn $\Delta\rho(T)$ depend on the concentration of defects. The “Matthiessen like” rule and especially the robustness of the high *T* variation of $\rho_n(T)$ with increasing defect content are totally unexpected observations.

*(b) Resistivity of the "pure" samples*

In $YBCO_{6.6}$ the *S* shaped contribution to $\rho_{pn}(T)$ is linked with the existence of the pseudogap, and application of fields as large as 60T were not sufficient to suppress SC, except when $T_c$ has been decreased by irradiation. The inflection point at high *T* has been found to occur approximately at $T^*/2$ and is seen to be independent of defect content. This confirms the independence of the PG on disorder established by NMR for Zn substitution as recalled in Fig.1, but also points out that the hole carrier content $n_h$ in the $CuO_2$ planes is not modified by the irradiation damage, at least in the high *T* range.

This conclusion is therefore as well valid for the slightly overdoped $YBCO_7$ inasmuch as the *T* linear strange metal behavior is independent of disorder. Similarly in the overdoped regime of Tl2201 the variation of $\rho_{pn}(T)=bT^p$ *with* $p\sim1.5$ applies rather well, which supports an approach towards a Fermi liquid behavior.

In all three cases as will be shown in **D**, a 60T field is sufficient to suppress the SCF contributions to the conductivity but does not allow a direct determination of $\rho_{pn}(T)$ below $T_c$ down to $T=0$. In some cuprate families with low optimal $T_c$ the linearity with *T* apparently extends down to rather low *T* without applied field. We found that linear fits of our high *T* data in our best pure $YBCO_7$ samples would imply a negative value of $\rho_{pn}(0)$. The *T* linear dependence of the strange metal behavior would therefore break down at low *T.* For the Tl2201 extending the $bT^p$ fit at low *T* would be compatible with $\rho_{pn}(0)=0$. On the contrary for $YBCO_{6.6}$ one does get a residual resistivity of about $\rho_{pn}(0)=1k\Omega/\square$ on Fig. 4b for the "pure" sample.

*(c) Metal insulator crossover?*

We have taken data in large enough applied field to suppress the SCF contribution to the conductivity for many defect contents. After subtracting the measured or extrapolated $\rho_{pn}(T)$ values we have determined the $\rho_0$ and $\Delta\rho(T)$ values defined in Equ. [1]. As $\rho_0$ was found to be a good measure of the defect content the ratio $\Delta\rho(T)/\rho_0$ has been reported versus $T$ in Fig. 5a (12).

For the underdoped sample one can see that a metallic behavior persists for the lower defect contents A and B. When the SC state has been nearly completely suppressed (C and D) the data still suggests an insulating state. For the optimally doped $YBCO_7$ this procedure was not applicable (46) except after the final irradiation that lowered $T_c$ to 1.9K. In that case a metallic behavior persists, as seen in Fig. 5a.

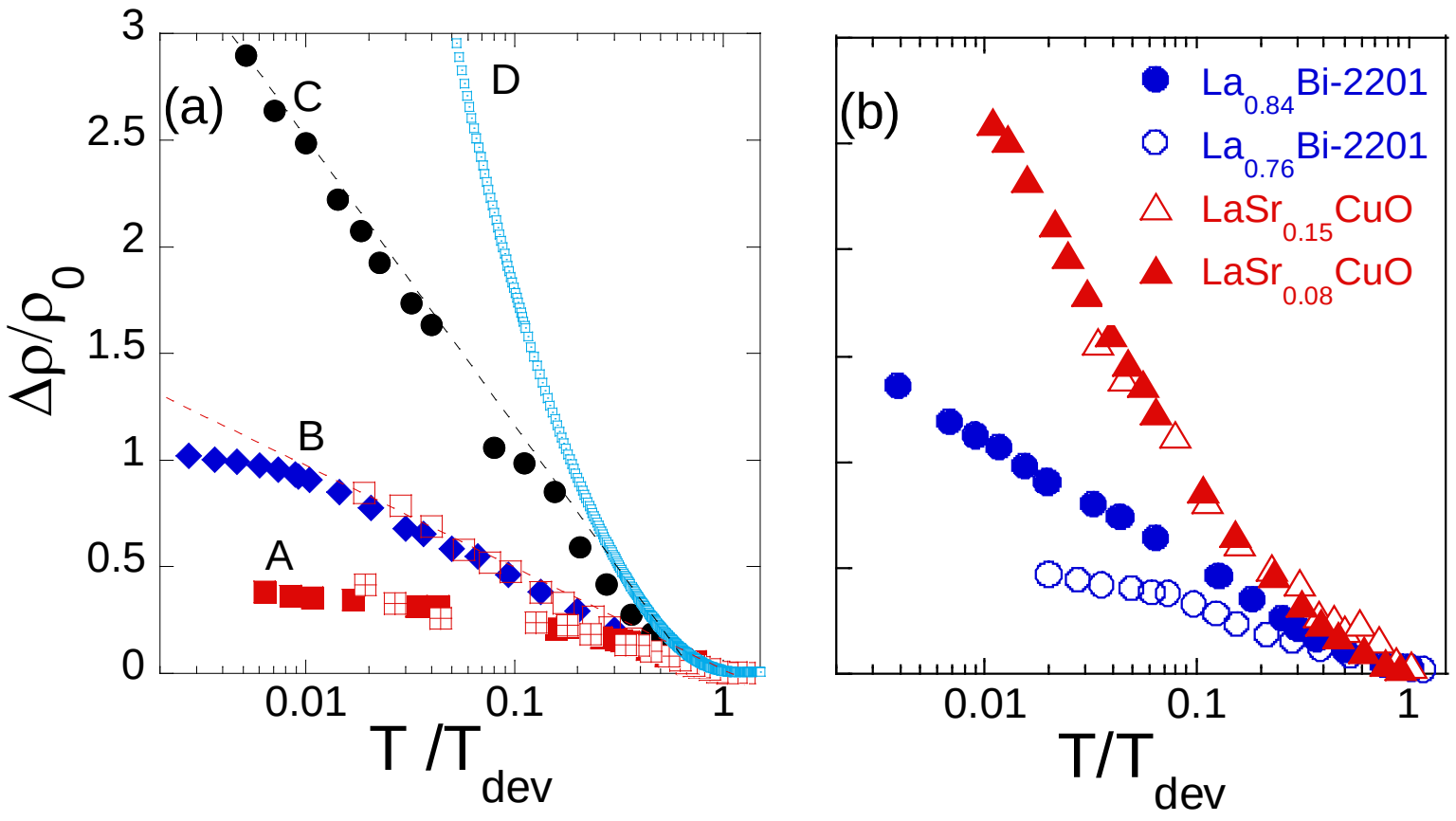

Fig.5: Low $T$ upturn $\Delta\rho$ normalized by the increase of residual resistivity $\rho_0$ plotted versus $T$ on a log scale. To permit quantitative comparisons $T$ is refereed to $T_{dev}$, the onset $T$ of $\Delta\rho$ whose values are between 100 and 150K (12). **(a)** for $YBCO_{6.6}$ (A to D) and $YBCO_7$ ( blue diamonds) **(b)** Similarly analyzed data for (LaSr)124 and (LaBi)2201 that are cuprates with low optimal $T_c$ (see text and ref. (12).

We have similarly analyzed some published data taken on low $T_c$ compounds (LaSr)124 (48) (49)and (LaBi)2201 (50). Those presented in Fig. 5b provide evidence that the resistivity upturns for these “pure” samples are quite similar to those found for YBCO after heavy electron irradiation. This analogy ascertains that a large disorder affects the physical properties even in the optimally doped (LaSr)124 family (49), as anticipated from the Phase Diagrams in section **B**.

Let us further note that in Fig. 5a the "insulating" behavior in the PG regime occurs only for very high in plane defect contents exceeding 5% per Cu. The disorder is then indeed at the origin of the apparent Metal Insulator Crossover (MIC). Contrary to conclusions taken from data on some low $T_c$ cuprates (49), large applied fields do not induce the MIC. They only usefully suppress SC and reveal the underlying MIC associated with the specific intrinsic or extrinsic disorder.

*(d) Resistivity upturns for low defect content*

We have evidenced that sizable resistivity upturns revealed by the applied field remain however for 1.6% defect content (curve A). In that case $\Delta\rho(T)$ onsets at $T_{dev}$ of about 100K and saturates at low $T$. An enlarged presentation of this low defect content regime is displayed in Fig. 6. There one can see that $\Delta\rho(T)/\rho_0$ is reproducible for distinct samples, increases initially as -log$T$ and appears nearly independent of defect content. This trend toward single impurity scattering behavior resembles the initial observations done in dilute Cu-Mn alloys that revealed the occurrence of Kondo effect in simple metallic states (51).

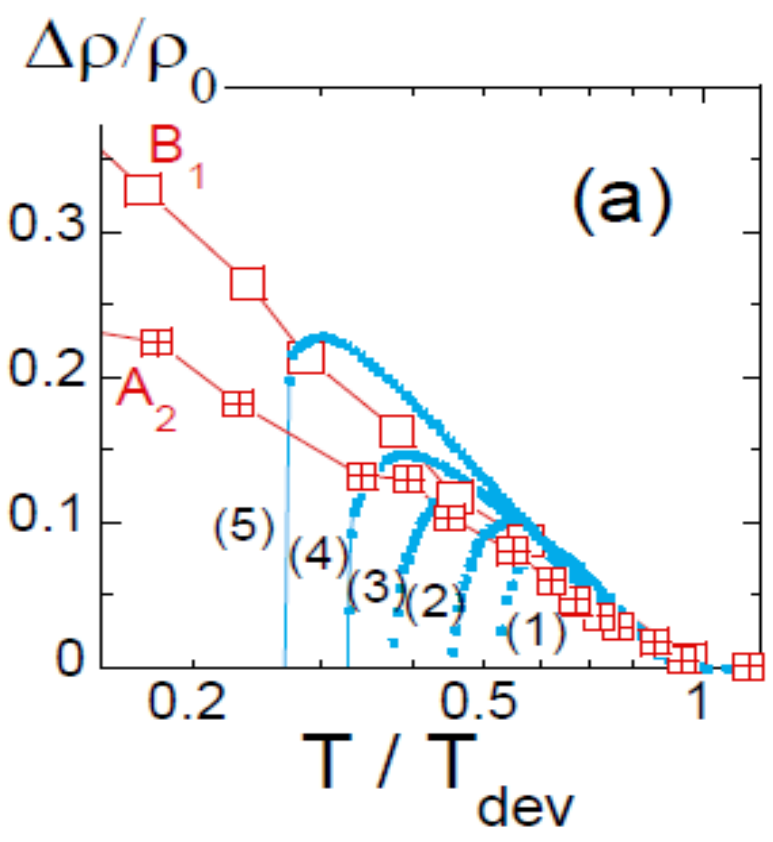

Fig. 6: Detailed high $T$ behavior of $\Delta\rho(T)/\rho_0$ of Fig. 5b taken for moderate point defect content on three distinct $YBCO_{6.6}$ samples. The five small blue squares curves are zero field data taken in situ for the same sample with increasing vacancy content from (1) to (5). Those for samples B1 and A2 taken in 60T pulse field allowed to suppress the contribution of SCF to the conductivity (see section **D**). From ref. (12).

The $^7$Li NMR experiments gave evidence for a paramagnetic response with a very low Curie Weiss $\Theta$ value that also resemble Kondo effect for the magnetic response. These similar findings on the transport and paramagnetism give a strong weight to this analogy.

The absence of variation of $T^*$ with defect content has led us to conclude that the doping $n_h$ is not modified upon introduction of disorder. We might assume that $n_h$ is as well $T$ independent. In a single carrier band model with $\rho_n(T)=m^*(n_h e^2 \tau)^{-1}$ the $T$ dependence would exclusively be due to the carrier's scattering rate. With this assumption the experimental results imply that the data involve two contributions $\tau^{-1}= \tau_{el}^{-1} + \tau_{in}^{-1}$. The elastic rate $\tau_{el}^{-1}$ responsible for $\rho_0$ scales with the defect content. The inelastic rate $\tau_{in}^{1}$ would be the sum of the inelastic rate of the pure compound $\tau_{in,p}^{-1}$ and of the Kondo like $\Delta\rho(T)$ contribution that also initially scales with defect content.

One does expects changes in the latter contribution for large defect content, due to the modification of their nanostructure. The interactions between impurities yield changes of scattering rates. Our estimate of a correlation length $\xi_{imp}$ of five lattice constants at low $T$ in $YBCO_{6.6}$ would indicate that a departure from the isolated defect regime should already occur for about a 1% vacancy content.

Similarly in $YBCO_7$ an upturn of $\rho(T)$ is only seen in Fig. 4a for an estimated defect content of about 8% . Even if $\xi_{imp}$ is about two lattice constants one would expect strong interactions between the defects. From the observed saturation at low $T$ of $\Delta\rho(T)/\rho_0$ seen in Fig. 5a an equivalent $\Theta$ value would be about 10K, much smaller than that of 150K found by $^7$Li NMR for in the strange metal regime for lower Li concentration. For Kondo alloys such as CuFe clusters of Fe impurities acquire a lower Kondo temperature than that $T_K$=30K of isolated impurities (52). The large increase of the resistivity upturns for high defect content in $YBCO_7$ would again appear quite similar to the situation encountered in CuFe alloys.

In Fig. 4 the resistivity upturns are much larger in the YBCO samples than in the overdoped Tl2201 sample. A thorough analysis permitted us to evidence that a weak localization induced by purely elastic scattering explains quantitatively the resistivity upturns in that case (47). That is again a good indication that the electronic properties approach a Fermi liquid behavior for highly overdoped cuprates.

### *3) Conclusion on the Normal state properties*

We have been able to disclose the incidence of dilute in plane point defects in the three considered "clean" materials that allowed us to probe respectively the PG, the strange metal and the Fermi liquid regimes. The Curie-Weiss magnetic perturbation induced by spinless defects and the low *T* saturation of the carriers scattering behave quite similarly with Kondo effect in classical Fermi liquids. The sharp increase of the Kondo like temperature $\Theta$ through the pseudogap line is a striking generic feature of the cuprates.

For large concentrations of defects, their mutual interactions might induce a metal insulator crossover. We could enhance defect clustering and evidence the importance played by the defect morphology in driving the insulating state. The comparison with data taken on "pure" lower $T_c$ compounds demonstrates that the incipient out of plane disorder induces similar upturns of the resistivity and an apparent MIC. This confirms our anticipations performed from simple comparisons of the Phase Diagrams in section **B**.

Can we reduce the disorder effects to study the Metal insulator transition in the small doping part of the phase diagram? Within our study of in plane defects we have shown that this disorder is well quantified by the residual resistivity $\rho_0$, but we do not know as accurately the action of out of plane defects. The present results provide evidence that they play an overwhelming role for low doping. The organization of the dopants certainly drive as well the stabilization of the family specific low *T* competing phases in the PG regime.

## (D) INCIDENCE OF DISORDER ON THE SUPERCONDUCTING STATE

From the 3D Phase Diagrams proposed in section **B**, we have noticed the existence of a correlation between the optimum $T_c$ value and the morphology of the Ph.D. for the various cuprate families. It is then important to analyze the evolution with increasing disorder of the superconducting properties that are available in our data of Fig.4. Such data could give indications on the actual origin of the pair condensed state, still poorly understood. In these 2D metallic compounds with low superfluid density the superconducting fluctuations are expected to persist above $T_c$. It has even been suggested that superconducting pair formation could occur at $T^*$ while $T_c$ would be the onset of phase coherence between superconducting pairs. For me this did not appear to be the case as our early experiments with Zn substitutions revealed that $T^*$ is not affected by disorder.

I shall consider hereafter in §1 the evolution of $T_c$ with controlled disorder that can be followed here until $T_c$ =0. In §2 I shall address the evolution of pairing strength through a study of the $T$ range of the superconducting fluctuations. This will allow me to discuss if there is any incidence on the SC properties of the Kondo like induced magnetism or of the resistivity upturns in the metallic PG regime.

### 1) *Variation of $T_c$ with in plane disorder*

Magnetic point defects usually induce pair breaking in classical superconductors. Here $YBCO_7$ exhibits a strange metal behavior with nearly zero residual resistivity in its normal state. Electron irradiation allowed us a full suppression of superconductivity though a metallic state persists. As displayed in Fig. 7 we found that $T_c$ scales linearly with the residual resistivity that scales itself with $n_d$, the defect content (46). Such a behavior disagrees with the Abrikosov-Gorkov (AG) variation (53) expected for pair breaking that implies a sharp downturn of $T_c(n_d)$ before reaching $T_c$=0. We gave further evidence for the inadequacy of AG from the low magnitude of the $\Delta T_c/\rho_0$ slope and the observed variation of the width of the SC transition. The latter, being due to the inhomogeneous distribution of defects remains narrow and does not broaden when approaching $T_c$=0.

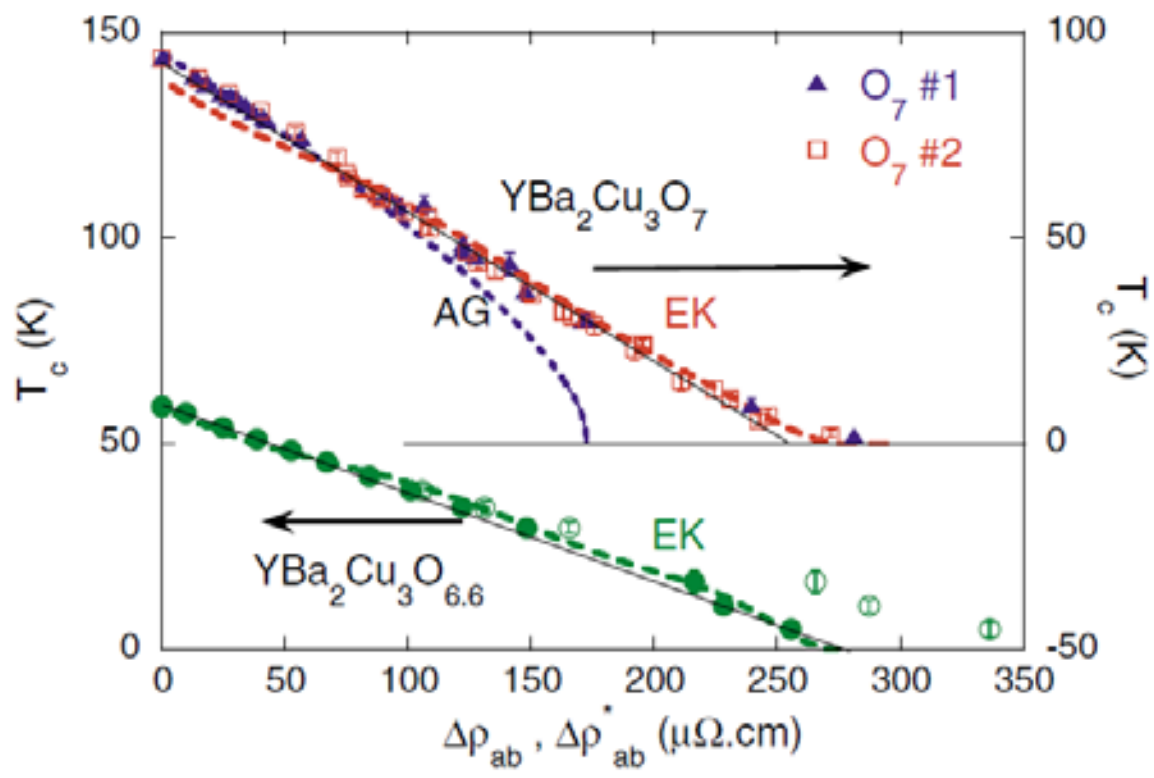

Fig. 7: Variation of $T_c$ with the excess residual resistivity $\rho_0$ induced by electron irradiation **(a)** for YBCO$_7$. Here $\rho_0 = \Delta\rho_{ab}$ is a measure of the defect content $n_d$. For comparison, the dotted lines report the expected pure pair breaking (AG) and phase fluctuation (EK) theoretical dependences. **(b)** For YBCO$_{6.6}$ the data are plotted for two different estimates of the defect content $\Delta\rho_{ab}$ (full circles) and $\Delta\rho_{ab}$* (empty circles).From ref. (46).

Emery and Kivelson (EK) have proposed (54) that a specificity of the cuprates is their weak superconducting carrier density. They suggested that in such a case phase fluctuations of the superconducting order parameter are at the origin of a $T_c$ reduction. The phase coherent SC transition should then occur when the resistivity reaches a critical value (55). This leads to a correlation between $T_c$ and $\rho(T_c)$ that better fits the experimental nearly linear defect dependence of Fig.7. A detailed discussion of the magnitude of $T_c$ done in ref (46) shows that the experimental situation probably requires to take both pair breaking and phase fluctuations into account. The incidence of phase fluctuations appears to be more important for large defect contents and in the underdoped PG regime, for which a quasi-linear variation of $T_c$ is also be observed in Fig. 7. This variation of $T_c$ with disorder already raised some questions. I shall show hereafter that more information is available from quantitative studies of the magnitude of the SCF.

2) ***Superconducting fluctuations and disorder***

Nernst effect (56), diamagnetism (57) (58) or paraconductivity (59) (60) measurements have been used to study the SCF above $T_c$. The main difficulty in such experiments resides in subtracting the normal state or spurious contributions from the data. Nernst effect data involve vortex motion contributions while spurious paramagnetic phases can counterbalance the diamagnetic signal.

*(a) Paraconductivity experiments*

We have rather preferred to use magnetoresistance (MR) measurements. As shown first in ref. (59) that permits us to obtain accurately the paraconductivity. Using 60T pulse fields on an $YBCO_{6.6}$ sample we demonstrated that the $H^2$ field dependence of the normal state MR known at high $T$ extends at temperatures well below $T_c$.

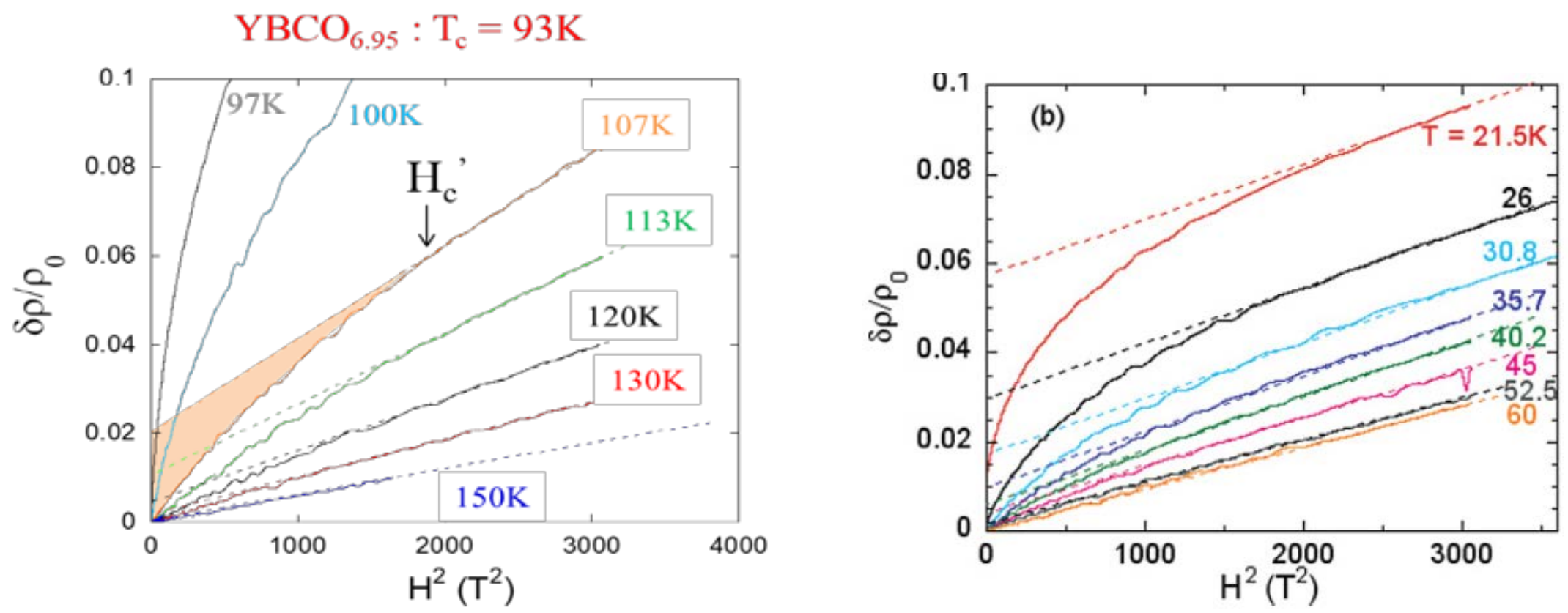

Fig. 8: Plot of the magnetoresistance *[ρ(H)- ρ (0) ]/ ρ (0)* versus $H^2$ at various temperatures **(a)** for pure $YBCO_{6.95}$. Here the orange area indicates the SCF contribution to the conductivity. **(b)** Data taken on $YBCO_{6.6}$ after decreasing $T_c$ to 5K by electron irradiation. Here the SCF appear below an onset of $T'_c$= 60K and are suppressed at fields $H'_c\ (T)$ that increase with decreasing $T$. From ref. (60).

Low field departures from the $H^2$ dependence progressively appear for decreasing $T$, when SC fluctuations contribute to the low field conductivity, as shown in Fig. 8a for $YBCO_7$. Such departures appear below an onset temperature $T'_c$. At each $T$ the normal state $H^2$ variation is recovered at a field $H'_c(T)$ (60).

The increase of conductivity $\Delta\sigma(T)$ associated with the SCF in zero applied field is easily obtained from the decrease of $\rho(T)$ with respect to the extrapolated normal state behavior. Typical results for $\Delta\sigma(T)$ and $H'_c(T)$ are displayed respectively in Fig.9a and 9b for pure and irradiated YBCO samples. As shown there both $T'_c$ and $H'_c$ decrease after reduction of $T_c$ *by* electron irradiation. Though the critical field $H'_c(T)$ cannot be determined at low $T$ for most samples, we could evidence that $H'_c(0)$ can be estimated using the "usual" variation

$$H'_c(T)= H'_c(0)\,[1-(T/T'_c)^2], \qquad [2]$$

that fits the data for low $T'_c$ samples (see Fig. 9b and ref. (59)).

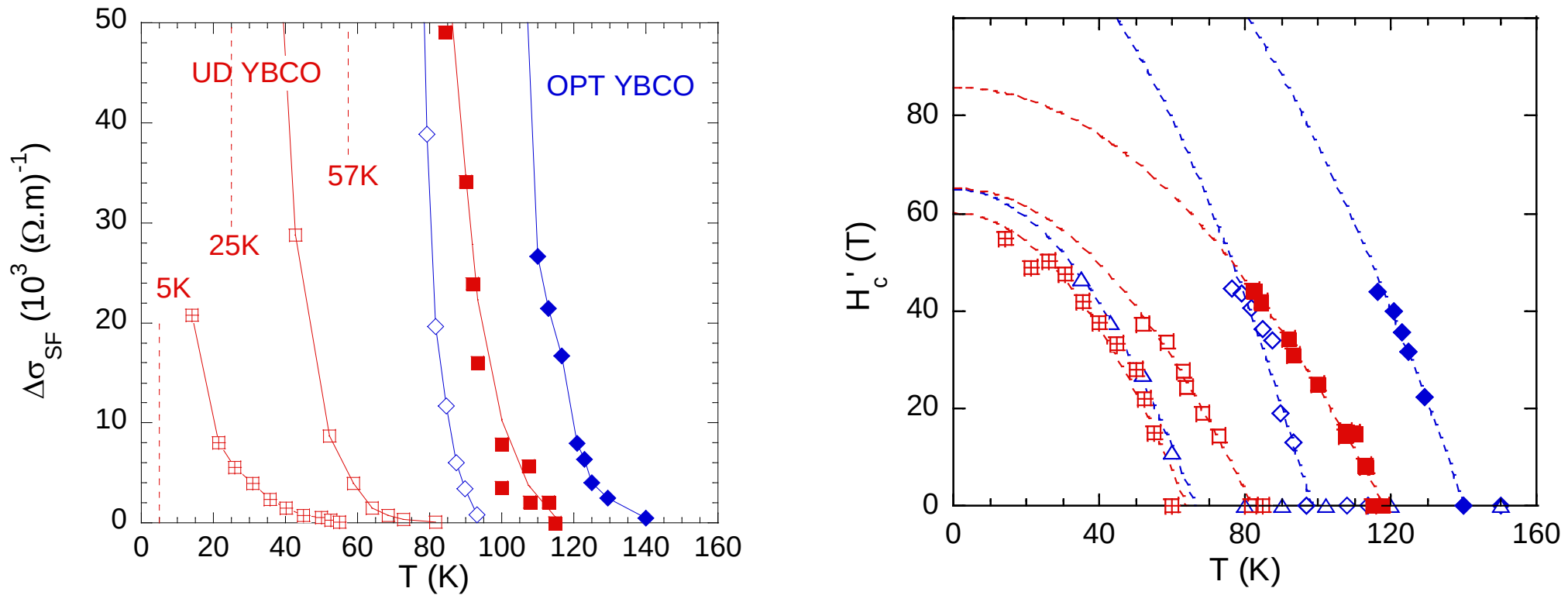

Fig. 9: **(a)** The zero field paraconductivity $\Delta\sigma(T)$ reported for pure and irradiated $YBCO_7$ and $YBCO_{6.6}$ vanishes for distinct $T'_c$ values. **(b)** $H'_c(T)$ data for the same samples. $H'_c(0)$ is estimated from the parabolic fits shown. From ref. (60).

*(b) SCF range and disorder*

The onset value $T_\nu$ of the SCF that we obtained on the same samples by Nernst effect (61) were smaller than the $T'_c$ deduced from the paraconductivity. The onset values obtained by a given technique are somewhat determined by the experimental sensitivity. Here they would decrease by about 10 to 20K for one order of magnitude decrease in the cut-off sensitivity. To our knowledge, the $T'_c$ values given here are among the highest reported in YBCO. More importantly, our experimental approach benefits of a significant technical advantage. Using a single sample and identical criteria for all defect contents ensures the reliability in the $T'_c$ data comparisons.

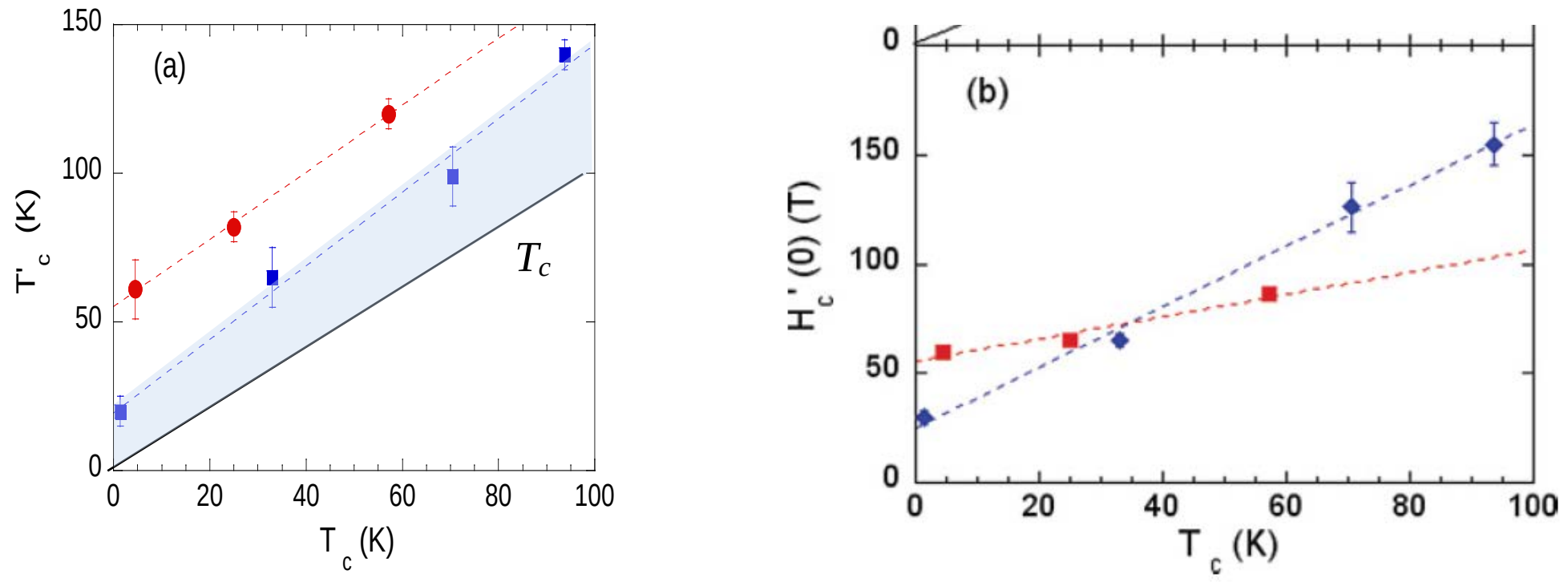

Fig. 10: Variation with in plane disorder of the SCF in the PG (red) and strange metal (blue) regimes. **(a)** Onset temperature $T'_c$ plotted versus $T_c$. The blue range is the SCF range for the $YBCO_7$ sample **(b)** Critical field value $H'_c(0)$ plotted versus $T_c$. It is less influenced by disorder for underdoped $YBCO_{6.6}$. From ref. (60).

We report in Fig.10 the results for $T'_c$ and $H'_c(0)$ versus $T_c$ obtained from the data of Fig. 9 for the $YBCO_7$ and $YBCO_{6.6}$ "pure" samples and after electron irradiation. One can easily notice that the SCF temperature range is quite similar for the different samples, but does not scale with the actual $T_c$ values. The ratio $T'_c/T_c$ increases steadily with defect content so that the range of SCF increases markedly with respect to $T_c$ in samples with in plane disorder. In the underdoped PG regime $T'_c$ remains as large as 60K for the $YBCO_{6.6}$ sample with $T_c$ =5K, and a large ratio $T'_c/T_c$ ~14 persists for the $YBCO_7$ sample in the strange metal regime for $T_c$ =1.5K.

From these simple observations, it appears quite clearly that the increase of in plane disorder does not reduce $T'_c$ as much as $T_c$. The fact that SCF remain important even if $T_c$ is highly depressed implies that superconducting pairs remain above $T_c$ while full 3D SC is not established. That suggests that phase fluctuations become more important in presence of disorder as suggested by EK in “bad metals” (55). A full quantitative study of the magnitude of the SCF conductivity helps to better clarify that point.

*(c) Amplitude and phase fluctuations in YBCO*

In an applied field $H$ the SCF paraconductivity is obtained by subtracting the normal state conductivity $\Delta\sigma_{SF}(T,H)=\rho^{-1}(T,H)-\rho_n^{-1}(T,H)$. Assuming that YBCO samples were among the cleanest cuprates, we have studied the $T$ and $H$ dependences of $\Delta\sigma_{SF}$ in ref. (60) for four samples from the PG to the overdoped regime. We have analyzed first the zero field data reported in Fig. 9a in the Ginzbourg-Landau framework. The SCF is associated with pair braking in that approach and the paraconductivity monitors the $T$ dependence of the superconducting coherence length

$$\xi_s(T)=\xi_s(0)/\varepsilon^{1/2} , \qquad [3]$$

with $\varepsilon= ln(T/T_c) \sim(T- T_c)/T_c$ for $T>T_c$. This 3D paraconductivity should diverge at $T_c$ on a very small $T$ range inaccessible experimentally, that evolves for a 2D compound towards a 2D Aslamazov Larkin (AL) regime (62) for which

$$\sigma_{SF}(T,0) = e^2/(16\hbar s\,\varepsilon), \qquad [4]$$

where $s$ is the interlayer spacing. The AL expression has been extended by Laurence Doniach (LD) (63) to include a $c$ direction coherence length $\xi_c(0)$. One can see in Fig. 11a that, assuming $\xi_c(0)$=0.9 Å, the LD analysis applies for $1.01<T/T_c<1.1$ for three of the YBCO samples. The $T$ dependence of $\Delta\sigma_{SF}(T,0)$ for the $YBCO_{6.6}$ sample is shifted to the right in Fig.11a but coincides with the other if $T_c$ is replaced by $T_{c0}$ =72K. I point out altogether here that in ref. (59) the normal state resistivity displays a $T$ dependence with a non-zero residual resistivity $\rho_0$ for that "pure" sample, as can be seen also in Fig. 4b. Considering the linear relation between $T_c$ and $\Delta\rho_{ab}$ of Fig.7 this $\rho_0$ value corresponds to a 10K decrease of $T_c$, so that a defect free sample could indeed have a critical temperature ~ $T_{c0}$.

As underlined in ref. (59), these data allowed us to point out that the existing uncontrolled disorder leads to an intermediate SCF regime between $T_c$ and ~$T_{c0}$. Phase fluctuations initially suppress $T_c$ in that $T$ range, while amplitude fluctuations of the superconducting order parameter take over beyond ~ $T_{c0}$.

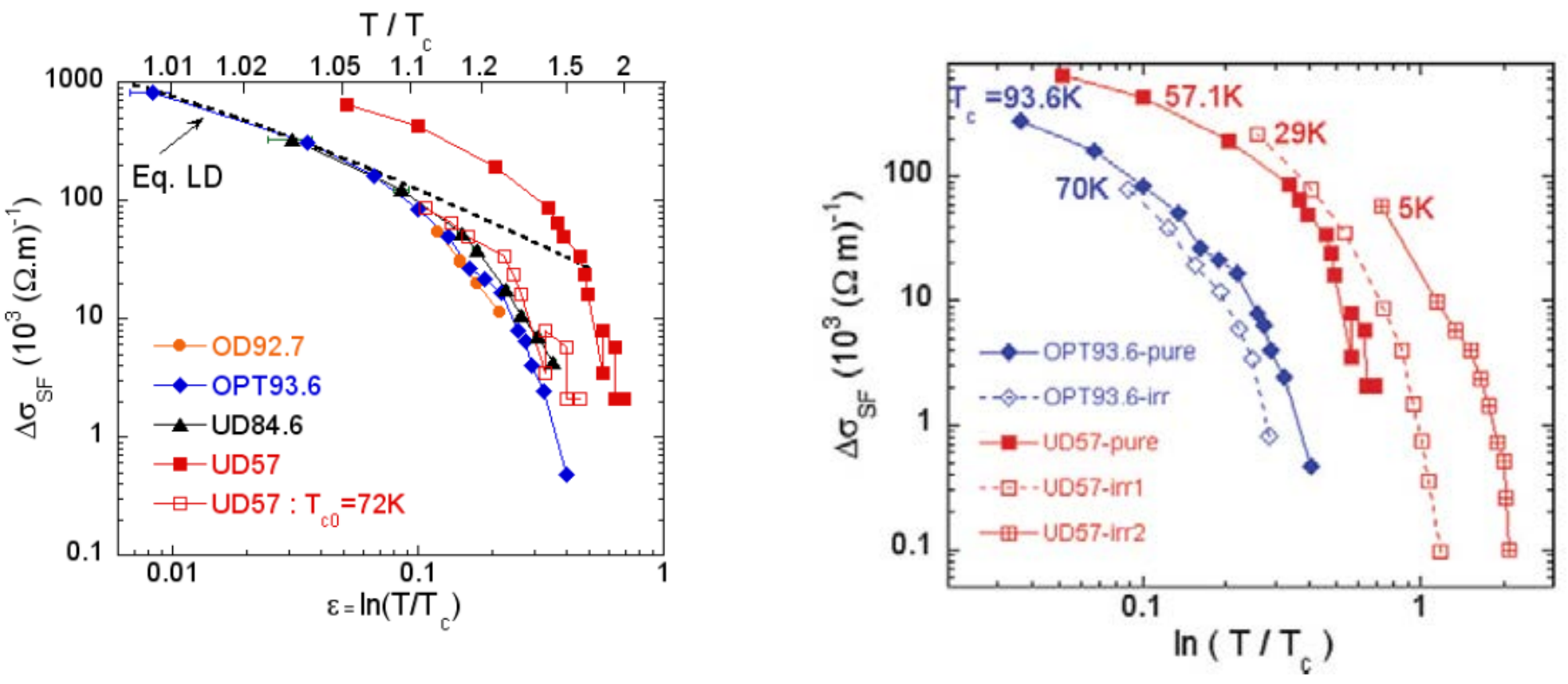

Fig. 11 **(a)** The paraconductivity $\sigma_{SF}(T,0)$ is plotted versus $\varepsilon = \ln(T/T_c)$ for the four "pure" YBCO samples with the indicated $T_c$ values. The dashed line represents the expected variation for the LD Gaussian fluctuations assuming $\xi_c(0)$=0.9 Å. For the underdoped sample UD57 with $T_c$=57K, the data are also plotted by replacing $T_c$ by $T_{c0}$ =72K (see text). **(b)** Similar plots for pure optimal (OPT) and underdoped (UD) samples and after electron irradiation that reduced $T_c$ to 70K for the OPT sample and to 29 and 5K for the UD sample. From ref (60)

The similar plots shown in Fig. 11b for the underdoped UD irradiated samples give evidence that systematic shifts to the right occur and increase with increasing defect content. Here again one could use values of $T_{c0} > T_c$ to get rough estimates of the increasing $T$ range for which phase fluctuations are dominant. For the optimally doped sample the $T$ variation of the SCF does follow the same pair breaking behavior as for the pure sample even though $T_c$ has been reduced by 23K.

These results indicate in the "pure" $YBCO_{6.6}$ sample the small $T$ range of phase fluctuation is presumably due to the residual incipient disorder. It broadens markedly with increasing in plane defect content. The GL amplitude fluctuations only become dominant at high temperatures in this PG regime. A totally coherent conclusion has been drawn from a quantitative comparison of the Nernst effect and $\sigma_{SF}(T,0)$ data, as detailed in ref (60). It fully supports a disorder induced large phase fluctuations range in the PG regime and the applicability of Gaussian fluctuations for the strange metal regime even in presence of disorder.

Let us consider now the case of the clean samples beyond the applicability of the LD amplitude fluctuations regime that applies up to $1.1T_c$ in Fig. 11a. Beyond that temperature $\sigma_{SF}(T,0)$ decreases markedly and vanishes quite sharply at $T'_c$ that might be assigned to a cutoff in the pairing. The latter might be linked with a minimal superconducting coherence length as discussed in (60).

We are still lacking a full theoretical approach in which the incidence of disorder on both phase and amplitude fluctuations would be treated on an equal footing. One can anticipate that the interplay between these two depends on the microscopic sources of superconductivity and disorder. But the present qualitative experimental information implies that $T'_c$ is a sharp temperature beyond which pair formation is energetically prohibited. $T'_c$ appears therefore as one of the best determination of the upper limit of the superconducting regime.

*(d) Magnetoconductance and superconducting gap*

Another important thermodynamic entity that characterizes the superconducting state is the field $H'_c(T)$ beyond which the SCF are suppressed. The highest measurable value of $H'_c$ is so far about 45T in the experimentally available 60T applied pulse fields. We have proposed that a natural extension down to $T$=0 may apply using Equ.(4) as usually done below $T_c$ for the $H_{c2}$ of classical 3D superconductors. This choice is supported by the data of Fig.9b taken on an OD sample with $T_c$ reduced to 5K by electron irradiation, for which we obtained $H'_c(0)$= 60T .

In the LD Gaussian fluctuation analysis one expects an independent determination of $H_{c2}$ from the low field behavior of the magnetoconductance $\Delta\sigma_H(T,H)= \sigma_{SF}(T,H)- \sigma_{SF}(T,0)$. We have reported in ref (60) an analysis of such data that surprisingly establishes that $H_{c2}(0)$ and $H'_c(0)$ are identical within experimental accuracy for the four pure samples considered in ref. (60) .

For large applied field $\Delta\sigma_H(T,H)$ is also found to depart from the LD behavior and displays then a sharp $exp[-(H/H_0)^2]$ decay. This provides experimental evidence that $H'_c(T)$ delineates a field beyond which pair formation is inhibited in analogy with the finding done for $T'_c$. This conclusion is valid even in the disordered cases, and the validity of this evaluation of $H'_c(0)$ is a further confirmation of the reliability of the extrapolation performed with Equ.(4) for the evaluation of $H'_c(T)$.

The important conclusion drawn from this detailed analysis is that $H_{c2}(0)$ or equivalently $H'_c(0)$ increase with doping. $H_{c2}(0)$ being a measure of $\xi(0)^{-2}$ one finds that the superconducting gap $\Delta_{SC}$ , which scales with $\xi(0)^{-1}$ also increases with hole doping. From the $H'_c(0)$ data displayed in Fig. 10b the incidence of disorder on the gap is found much smaller for the underdoped sample than for the optimally doped one. This appears in accordance with the corresponding data for $T'_c$ on Fig. 10a and with a dominance of phase fluctuations in the PG regime.

### 3) **Disorder in the 3D YBCO Phase diagram**

Though we have clearly shown that the underdoped $YBCO_{6.6}$ sample is not a perfectly clean cuprate, we have still confirmed that the YBCO family is not far from being clean enough to yield important conclusions on the differences between the PG and strange metal regimes. The detailed investigation of the SCF done in ref. (60) and reported here allowed us to determine reliably the temperature $T'_c$ beyond which superconducting pairing is prohibited, as the coherence length becomes presumably too small. As seen in Fig.12a $T'_c$ only displays a small decrease below optimal doping but follows $T_c$ and not the pseudogap $T^*$. We have also shown that the latter crosses the $T'_c$ versus doping line (Fig. 12a). This again confirms that the pseudogap is not a precursor to SC, as anticipated from the Zn experiments.

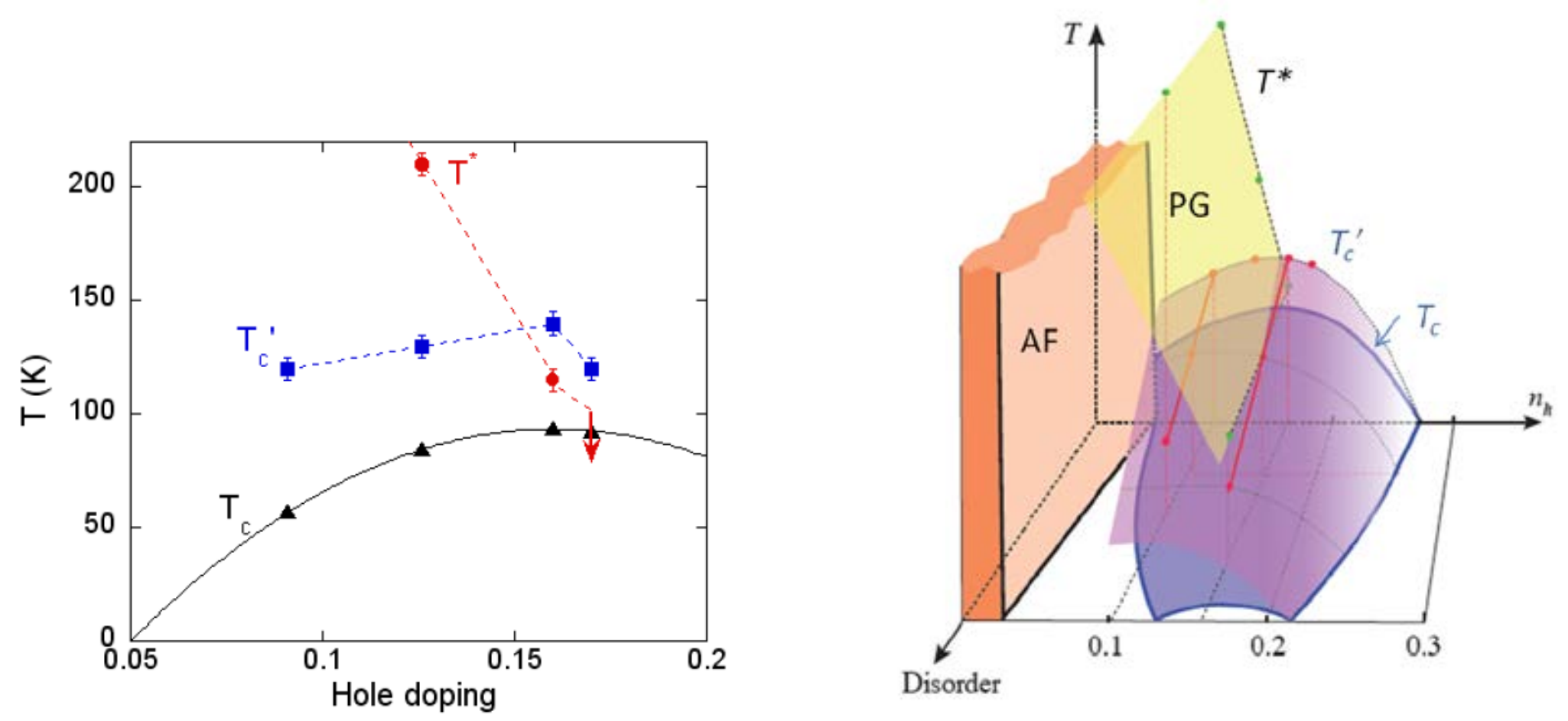

Fig.12 **(a)** Doping dependence of the 3D superconductivity $T_c$, the pseudogap $T^*$, and the SCF onset $T'_c$ for four "pure" YBCO samples **(b)** 3D phase diagram displaying the evolution of these temperatures with the concentration of in plane defects. The red lines are actual data for $T'_c$ of underdoped and optimally doped YBCO. The SCF remain large even when $T_c$ vanishes. From ref. (60).

The SC gap as determined by $H'_c$ from thermodynamic arguments increases steadily with hole content through the optimal doping. This appears in line with the doping dependence of the small gap observed by some spectroscopic experiments in other cuprate families (64) (65).

Our studies of the SCF in the "pure" samples also allowed us to point out that in the PG regime of $YBCO_{6.6}$ the incipient disorder slightly reduces $T_c$ that should be 10 to 15K higher in a "cleaner" sample.

The detailed investigations with increasing controlled disorder allows us by analogy with Fig. 2 to propose the 3D phase diagram of Fig. 12b specific to YBCO with in plane disorder. Here the disorder axis is better monitored than in Fig. 2, using the concentration of defects introduced by electron irradiation or equivalently the increase of residual resistivity $\rho_0$, for instance at optimal doping. In this phase diagram one should consider that $T'_c$ is a better identifier of the 2D pairing than $T_c$ that marks the 3D coherent superconducting state.

The analysis of the SCF done in ref. (60) and shortly reviewed here has further allowed us to differentiate the incidence of disorder on the SCF in the PG and strange metal regimes. In the latter the disorder induces essentially a pair breaking so that $T'_c$, $T_c$ and $H'_c$ decrease markedly altogether as seen in Fig 10. In the PG regime of underdoped $YBCO_{6.6}$ a large $T_c$ decrease is observed while $T'_c$ and $H'_c$ are only moderately reduced. This indicates that pairing remains while a loss of phase coherence suppresses 3D superconductivity. The disorder induces in the PG phase a large SCF range that extends from the reduced $T_c$ up to $T'_c$.

These conclusions about the phase diagram in the PG to strange metal regime are quite important as they result from experimental studies on the YBCO family that is a clean cuprate family as discussed in section **B**. For lower hole content $n_h$<0.1 the 3D $T_c$ decreases until the interplay between SC and AF occurs as displayed in Fig.1. In that range the doping $n_h$ is however not easy to control as it depends on the formation in the intermediate $Cu^+$ layer of $Cu^{2+}$ - $O^{2-}$ - $Cu^{2+}$ segments that do not produce any charge transfer in the $CuO_2$ planes. The Metal Insulator Transition could only be investigated with significant incipient out of plane disorder in $YBCO_{6+x}$ for x<0.45 or in $YBCO_6$ with $Ca^{2+}$ substitution on the $Y^{3+}$ site. Unfortunately the investigation of the overdoped range cannot be performed either in the YBCO family, as it can only be reached by $Ca^{2+}$ substitution that also introduces a sizable disorder. In Fig. 12b we have reported in that range a $T_c$ variation analogous to that observed for Tl 2201 and assumed that amplitude fluctuations remain dominant to anticipate the $T'_c$ variation.

## (E) DISCUSSION

I shall summarize first hereafter the various results concerning the phase diagram and underline the experimental questions that remain open. I shall analyze then the important information disclosed by using controlled defects as probes of the original properties of the distinct states of the phase diagram.

***Phase diagrams and Low $T_c$ cuprates***. We have done systematic investigations of the evolution of the cuprate phase diagrams with increasing controlled disorder in the $CuO_2$ planes. Our initial Zn substitutions experiments demonstrated that the disorder opens a large spinglass regime as displayed in the 3D phase diagram of Fg. 2. The controlled increase of vacancy content in single crystals allowed us then to evidence that SCF persist above $T_c$ up to a temperature $T'_c$ that decreases less than $T_c$ with increasing disorder. The normal state residual resistivity $\rho_0$ increases markedly altogether and low $T$ upturns appear mostly in the underdoped regime for large disorder.

All these features had been observed in the "pure" singular low $T_c$ cuprate families such as LSCO (48) or (LaBi)2201 (50). Large residual resistivity $\rho_0$ were clearly apparent in the data for those cuprate families, for which the large fields applied to suppress $T_c$ disclosed resistivity upturns even at optimum $T_c$ (49). Nersnt effect data revealed as well a large range of SCF above $T_c$ in the underdoped rPG regime (56). All these observations confirm our initial guess that had led us to propose the 3D Phase diagram of Fig. 2, and allowed us to anticipate the presence of a large incipient disorder that diminishes $T_c$ in those specific cuprate families.

Comparing the $\rho_0$ of these low $T_c$ compounds with those induced by electron irradiation in our YBCO samples even suggests that the linear relation of Fig. 7 between $T_c$ and $\rho_0$ might have an extended validity. Choosing then to use the $\rho_0$ value at optimal doping to fix the position of the Ph. D. of the diverse cuprate families on the disorder axis in Fig. 2 would not modify it significantly (66). In the phase fluctuation scenario a relationship between $T_c$ and $\rho$ $(T_c)$ is indeed expected. This ensemble of arguments therefore lead me to *conclude that the incipient disorder is mostly at the origin of the reduced $T_c$ values in these anomalous cuprate families.*

***Towards disorder free cuprates?*** Once we have considered that we should put aside the low $T_c$ cuprates, can we decide what family better represents clean cuprates? For the YBCO, Bi2212 , Hg1201 and Tl2201 that display the highest $T_c$~90K we noticed that the extrapolated $\rho_0$ is nearly zero at optimal doping (66). These families are therefore nearly equivalent in that doping range. We could indeed initially establish the unicity of the transition *T** between the PG and the strange metal regimes in YBCO and Hg1201 (11).

Our phase diagram of Fig. 2 give evidence that determination of the hole contents $n_h$ by mapping the SC "domes" between different cuprate families is certainly incorrect, except for the higher $T_c$ families. Unfortunately it appears difficult to underdope the Tl2201 family that is the only "clean" cuprate for which the overdoped to non-SC transition is accessible. As indicated in Section **D** we have not been able so far to perform a full study of the variation of the thermodynamic properties of the SC state through that transition.

$YBCO_{6+x}$ is similarly the only family on which the Metal to Insulator transition occurs without a spin glass intermediate regime. We have however seen here that for $x$=0.6 incipient disorder effects already appear as $T'_c$ remains large with respect to $T_c$. Low *T* upturns of the resistivity occur for oxygen contents $x$<0.4 (12), but we do not know how the insulator state sets in. We do not have any reliable control on the incidence of the oxygen disorder during this crossover.

One cannot anticipate whether a first order SC- insulator transition would occur in a disorder free cuprate, hence the question mark we have introduced in the 3D phase diagram in Fig. 2.

So although huge progresses have been performed since the discovery of the cuprates, we only have at hands a limited number of cuprates that allow to explore independently some characteristic clean compositions in the phase diagram of Fig. 2. We were not able to disclose so far a batch that would play for the cuprates a similar role to that played by Silicon single crystal wafers for semiconductors.

I should like to point out that the present critical concern is even more pertinent if one considers the actual disorder that occurs on the sample surfaces. The latter are investigated using novel techniques restricted to easily cleavable single crystals. ARPES (67) and STM (68) experiments allowed to thoroughly study the doping dependence of the **k** space surface band structures of Bi2212 crystals. One can beautifully see there the dopant disorder in the spatially resolved STM spectra (68) that have been helpful to correlate electronic structure with the local order. This unavoidable dopant incipient disorder probably limits the energy resolution in ARPES spectra. The bulk and surface dopant disorder are not necessarily identical so that refined comparisons between bulk thermodynamic properties and surface energy spectra might be misleading. Local STM and NMR experiments have been achieved on Zn substituted cuprates. NMR could give information mostly on the bulk YBCO normal state properties, while STM permitted to probe the SC local states on the Bi2212 surface (69).

***Point defects in the normal state.*** The change in the normal state transport properties have initially revealed the difference between the PG, the strange metal and the quasi Fermi liquid regimes in the "pure" cuprates. The introduction of point defects has allowed us to highlight those differences by providing clear evidence that the scattering of the carriers splits into two distinct contributions. *That observed in the pure compound is totally unmodified by the introduced defects.* It adds to the specific scattering of the carriers by the dilute impurities. An unexpected difference with the situation encountered in classical metals is that in the latter case the scatterings are associated with two different processes, an elastic scattering on the defects and an inelastic one mediated by phonons modes. In the cuprates both processes deal with electronic states controlled by the correlated electronic properties of the pure compound.

This stems from the fact highlighted by the NMR experiments that the impurity substitutions or vacancies in the $CuO_2$ planes induce a modification in their vicinity, which has a significant spatial extension that varies with temperature and doping (38). This is certainly an unambiguous evidence that the cuprates are strongly correlated electron systems.

We evidenced as well low *T* upturns of the resistivity when introducing in plane vacancies by electron irradiation. Those correspond to scattering rates on the defects similar to those observed for Kondo magnetic impurities in classical metals. As recalled in section **C** these upturns scale with defect content as long as their concentration is small enough to avoid overlaps between the clouds induced by the vacancies.

We have shown, both by NMR and transport, that a sharp modification of the defect state occurs when the hole doping of YBCO is switched from the PG to the strange metal regime. We assign that to a reduction of the paramagnetism similar to a large increase of the Kondo like temperature. In the strange metal regime the cloud size does not exceed two lattice constants and the resistivity upturns decrease, in agreement with an increased “Kondo” temperature. We assume that the paramagnetism should disappear when reaching the pure Fermi liquid regime for large hole doping. This crossover remains however to be evidenced for instance on the Tl2201 cuprate family.

***Disorder in the SC state***. The data analysis of section **D** reports an original study of the destruction of the 2D superconductivity by a controlled disorder. The GL or Lawrence-Doniach pair-breaking approach explain rather well the 2D SCF up to 1.1 or 1.2$T_c$ in the pure or disordered YBCO near optimal doping. That would apparently also hold for very clean underdoped samples.

In that PG regime, any disorder incipient or produced by electron irradiation, induces a large SCF range as compared to $T_c$, that mostly originates from a loss of phase coherence (60). Emery and Kivelson (55) had proposed a major importance of phase fluctuations for SC with either a low superfluid density or a large resistivity at $T_c$. Our study indicates that the disorder is mainly responsible for the phase fluctuations in underdoped YBCO. I therefore suggest that the Kondo like inelastic defect scattering of the hole carriers is the main player in the $T_c$ reduction that occurs before the SC -insulator transition.

One should notice finally that our study of the field dependence of the SCF give a measure of $H'_c(0)$, the critical 2D magnetic field. Thermodynamic arguments allow us to connect it with $\Delta_{SC}$, the SC gap magnitude. In clean YBCO $\Delta_{SC}$ increases with doping through the optimum $T_c$ and correspondingly the coherence length $\xi(0)$ diminishes. One expects to reach a cutoff value in the overdoped regime that might be probed experimentally in the Tl2201 family.

## (F) CONCLUSION

We have recalled here that the cuprates are correlated electron systems in which incipient disorder occurs naturally due to the chemical doping. This disorder has introduced great difficulties in the understanding of the very phase diagram of the hypothetically clean cuprate plane. Immediately after the discovery of YBCO we have highlighted that the pseudogap to strange metal transition is robust to disorder at high temperatures. We have evidenced that it is generic, which means identical for all hole doped cuprate families. A long time has been required though (25) to see that fact accepted by the community.

A large attention has been devoted to the original and interesting ground states discovered, such as stripes in the 124 family and later on CDW in YBCO. This diversity of ground states in the cuprate families testifies the richness of the underlying physics of these correlated electron systems. Meanwhile we have been doing with collaborators an extensive work on $Na_xCoO_2$ (70) (71) that demonstrated that the physical properties of the $CoO_2$ layers were highly influenced in these compounds by their interactions with the Na dopants. Though similar detailed investigation haves not been achieved so far in the cuprates, I got convinced that there too the order of the dopants may analogously play a great role in stabilizing the different ground states. The fact that these states compete with SC complicated our understanding of the variety of cuprate phase diagrams.

As recalled here, we proposed the 3D phase diagram of Fig. 2 with some Orsay-Saclay colleagues. We suggested that the AF insulator, the PG, the strange metal, and the Fermi liquid metallic regimes were the actual phases that characterize the high *T* part of the disorder free cuprate phase diagram. Among those, all the metallic states display a superconducting ground state except for large doping in the Fermi liquid range. On experimental grounds the less understood part of the phase diagram is the transition from the metallic PG regime to the insulating state presumably because even the weak incipient disorder has a strong incidence there on the physical properties.

In my research group, we have always considered that the intentional introduction of controlled impurities or defects allowed us to disclose physical properties of the cleanest phases. The original "Kondo like" magnetic and transport responses induced by spinless impurities are strong signs for a connection with Mott physics. The step change of its characteristic temperature $\Theta$ has been somewhat overlooked so far by the community though it characterizes a generic part of the clean cuprate phase diagram.

This transition through $T^*$ recalled in Fig. 3b is so abrupt that it is legitimate to wonder whether it marks a phase transition rather than a crossover. Though $T^*$ is not modified by disorder it could still correspond to a transition that would only be slightly smeared by the weak incipient disorder always present in the cleanest compounds. The sharp drop of the $T^*$ line versus $n_h$ near optimal doping with a quantum critical point could agree with such a possibility. Whether orbital currents (72) (73) appear at $T^*$ remains an open question and one should like to understand how such a collective state would induce a Kondo like magnetic cloud around spinless defects. Any further experimental input on such a matter would certainly require extremely clean samples.

If I consider now the SC state, we have evidenced that the onset of 2D superconducting pair formation occurs rather sharply at a higher temperature $T'_c$ than the 3D superconductivity at $T_c$. Most researchers considered that the $T_c(n_h)$ dome shape is universal for the cuprates, with differing $T_c$ magnitudes. We oppositely evidenced that $T'_c$ is always large, even in the anomalous cuprate families for which the incipient disorder (or order) imposes a low $T_c$ value. As shown in Fig. 12a we have even demonstrated that one might hardly define a dome for the $T'_c\,(n_h)$ variation for the cleanest cuprates.

This study of the incidence of defects has therefore disclosed two apparently contradictory observations. The PG line marks a high $T$ transition between different electronic states, while the low T pairing energy is nearly unaffected by the opening of the pseudogap, which does not therefore compete directly with SC. Our SCF experiments with controlled disorder allowed us to evidence that the only difference in the SC states results from the incidence of in plane defects. They induce pair breaking in the strange metal while they furthermore enhance phase fluctuations in the PG regime. As recalled previously, we do not know whether $T'_c$ would decrease and vanish before reaching the SC to insulator transition in a “clean” cuprate.

The low carrier density and the associated low phase stiffness has been suggested as a possible source of phase fluctuations and $T_c$ reduction in cuprates (54). Our data rather imply that large impurity induced carrier scattering rates are required to promote a bad metal behavior. This leads me to speculate that the distinct incidences of disorder on SC could be due to the actual difference in the normal state Kondo like response. That would qualitatively be consistent with the decrease of the 3D $T_c$ on the left of the dome like SC regime in the supposedly pure cuprate families.

I noticed more than ten years ago (25) that the occurrence of the pseudogap and the SC states in the cuprates finds some theoretical justification in computations done by Cluster DMFT within the simple one band Hubbard model (74). Within the three band Emery model (75) similar computations give analogous results and introduce realistic extra parameters, such as the charge transfer gap or the hole occupancies of the oxygen and copper orbitals. Those are considered nowadays to try to justify the difference of optimum $T_c$ between the diverse cuprate families (76). Our work establishes that disorder is a dominant factor in this differentiation of the low $T_c$ cuprates, while $T_c$ only marginally differ for cleaner cuprate families with low residual resistivity. The actual incidence of the planar structure is therefore less important on the optimal $T_c$ than might be anticipated from the data. The numerical computations being done on clean models, it would appear quite interesting to see if the three-bands parameters have any influence on the SC dome shape for the clean compounds. Such computations could help to suggest a behavior for the SC –Insulator transition that is at present obscured by disorder in the real materials.

As for the relation between the normal and SC states properties, I hope that the present experimental results will stimulate extensions of the thermodynamic developments taking into account random disorder, amplitude and phase fluctuations on the same footing. This could permit hopefully to relate the normal state scattering properties with $T'_c$ or $T_c$ and to open a route toward an understanding of the microscopic origin of the pairing energy.

## ACKNOWLEDGEMENTS

I would like to acknowledge here may collaborators and students who have participated in the large set of experiments mentioned here. Florence Rluuler-Albenque (CEA Saclay) has been responsible for most experimental work on transport properties and would have been a co-author on this article if she did not pass away too early in 2016. P. Mendels has been an efficient collaborator on the NMR experiments done during many years after the cuprate discovery and J. Bobroff has done essential contributions for his PhD and well after with A. Mac Farlane and with his student S. Ouazi. Collaborations with G. Collin, J.F. Marucco (ICMO Orsay), N. Bamchard, D. Colson and A. Forget (CEA Saclay) have been essential for all the preparation and characterization of the NMR and single crystal samples. F. Balakirev initiated us to perform the first pulsed high field transport measurements achieved at the NHMFL in Los Alamos. We complemented them by many runs performed at the SNCI (Toulouse), with the help of B. Vignolles, D. Vignolles and C. Proust. Over the years, we have benefited of numerous exchanges on the experiments with K. Behnia, P. Bourges and Y. Sidis and on theoretical matters with M. Civelli, M. Gabay, A. Georges, M. Héritier, P. Hirschfeld, C. Pépin, H. Schulz, G. Sordi and A.M. Tremblay.